\documentclass[aps,prb,reprint,amsmath,showpacs,showkeys]{revtex4-1}

\usepackage{graphicx}
\usepackage{dcolumn}
\usepackage{bm}
\usepackage{float}
\begin{document}


\title{Heavy Fermion superconductor CeCu$_2$Si$_2$ under high pressure:\\ multiprobing the valence cross-over}



\author{G. Seyfarth}
\email[]{gabriel@seyfarth.net}
\affiliation{DPMC - Universit\'e de Gen\`eve, 24 Quai Ernest Ansermet, 1211 Geneva 4, Switzerland}
\author{A.-S. R\"uetschi}
\affiliation{DPMC - Universit\'e de Gen\`eve, 24 Quai Ernest Ansermet, 1211 Geneva 4, Switzerland}
\author{K. Sengupta}
\affiliation{DPMC - Universit\'e de Gen\`eve, 24 Quai Ernest Ansermet, 1211 Geneva 4, Switzerland}
\author{S. Watanabe}
\affiliation{Quantum Physics Section, Kyushu Institute of Technology, Fukuoka 804-8550, Japan}
\author{K. Miyake}
\affiliation{Graduate School of Engineering Science, Osaka University, Toyonaka 560-0043, Japan}
\author{A. Georges}
\altaffiliation{also: Centre de Physique Th\'eorique, \'Ecole Polytechnique,
  CNRS, 91128 Palaiseau Cedex, France}
\altaffiliation{Coll\`ege de France, 11 place Marcelin Berthelot, 75005
Paris, France}
\author{D. Jaccard}
\affiliation{DPMC - Universit\'e de Gen\`eve, 24 Quai Ernest Ansermet, 1211 Geneva 4, Switzerland}


\date{\today}

\begin{abstract}
The first heavy fermion superconductor CeCu$_2$Si$_2$ has not revealed all its striking mysteries yet. At high pressures, superconductivity is supposed to be mediated by valence fluctuations, in contrast to ambient pressure, where spin fluctuations most likely act as pairing glue. We have carried out a multiprobe (electric transport, thermopower, ac specific heat, Hall and Nernst effects) experiment up to $7~\text{GPa}$ on a high quality CeCu$_2$Si$_2$ single crystal. For the first time, the resistivity data allow quantitatively to draw the valence cross-over line within the $p-T$ plane, and to locate the critical end point at $4.5\pm0.2~\text{GPa}$ and a slightly negative temperature. In the same pressure region, remarkable features have also been detected in the other physical properties, presumably acting as further signatures of the Ce valence cross-over and the associated critical fluctuations: we observe maxima in the Hall and Nernst effects, and a sign-change and a strong sensitivity on magnetic field in the thermopower signal.
\end{abstract}


\pacs{74.70.Tx, 74.62.Fj, 74.25.fc, 74.25.fg}



\maketitle



%


\section{Introduction\label{}}

\begin{figure}[tb]
\includegraphics[width=\linewidth]{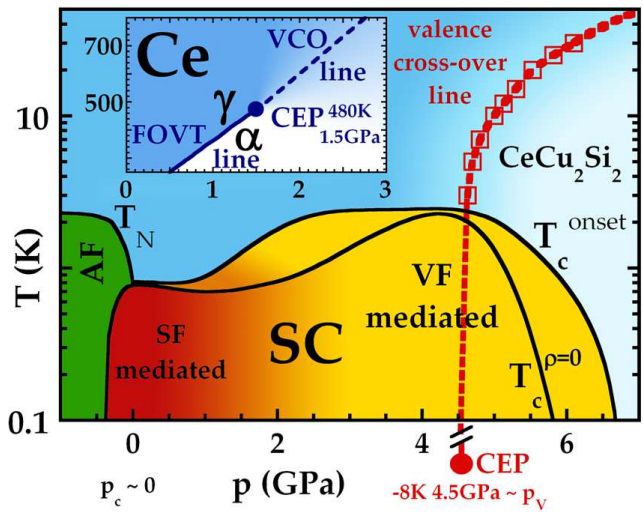}
\caption{Updated $p-T$ phase diagram of CeCu$_2$Si$_2$ with its superconducting phase extending far beyond the magnetic instability at $p_c$. The maximum in $T_c(p)$ roughly coincides with the position of the critical end point (CEP); the valence cross-over (VCO) line (red) is obtained from $\rho(p)$ data and corresponds to that in fig.~\ref{fig4}a. For negative $p$, data \cite{Didier_1999} on CeCu$_2$Ge$_2$ were used for a qualitative outline, and between 0 and 2~GPa the smooth interpolation is based on refs. \onlinecite{Vargoz_1998,Alex_thesis}. See section~\ref{sec_rho} for more details, particularly concerning the distinction between $T_c^{\text{onset}}$ and $T_c^{\rho=0}$. Inset: $p-T$ plane of elementary Ce in vicinity of the CEP of the first-order valence transition (FOVT) line (according to Lipp et al.\cite{Lipp_2008}).}
\label{fig1}
\end{figure}

Even though investigations started already in 1979 \cite{Steglich_1979}, CeCu$_2$Si$_2$ still carries key enigmas of heavy fermion (HF) superconductivity (SC). Numerous (pressure) studies on other Ce-based HF compounds, notably concerning the 115-family, focus on the magnetic instability or magnetic quantum critical point as a driving force for exotic behaviors such as unconventional SC. However, in reference to the case of elementary Ce, it is known that the Ce valence changes with pressure (delocalization of the $4f$ electron) and that this effect may also play a major role in the determination of electronic properties. As a consequence, observations in Ce compounds may result from a complex combination of different underlying microscopic phenomena. In this respect, CeCu$_2$Si$_2$ may be a good candidate to help to disentangle the contributions from the spin and charge degrees of freedom, more intricate in other Ce systems (like the 115-family). In particular, the pressure~($p$) - temperature~($T$) phase diagram (fig.~\ref{fig1}) of CeCu$_2$Si$_2$ deviates from the ``generic'' one in which superconductivity (SC) emerges rather close to a magnetic instability, presumably mediated by critical spin fluctuations (SF). Instead, in CeCu$_2$Si$_2$ the superconducting region extends far beyond the magnetic quantum critical point (located at $p_c\approx0$~GPa), and exhibits an enhanced transition temperature ($T_c$) culminating at $\sim2.4$~K around $4$~GPa \cite{Didier_1985,Didier_1999,Alex_PRB_2004,Alex_2007}. The most elaborate scenario \cite{Miyake_2007,Wata_2011} invokes critical charge or valence fluctuations (VF) associated with an underlying valence transition (VT), i.e. delocalization of the Ce $4f$ electron when tuning $p$ across the critical region ($p_V$). In elementary Ce (see inset of fig.~\ref{fig1}), the first-order valence transition (FOVT) line ($\gamma-\alpha$ transition), the location of its critical end point (CEP) and the valence cross-over (VCO) line have been unambiguously identified by several experiments \cite{Jayaraman_1965,Lipp_2008}. In CeCu$_2$Si$_2$, obviously the CEP lies at much lower $T$, where the electron delocalization is accompanied in the transport data by a decrease in resistivity, in combination with SC at the lowest $T$. Preceding experiments point towards distinct natures of the two sc phases \cite{Alex_PRB_2004,Miyake_2007} (related to $p_c$ and $p_V$ respectively), and have unveiled close to $p_V$ features peculiar to the concept of critical VF, such as a $T$-linear resistivity and an increased residual resistivity $\rho_0$ \cite{Didier_1999,Alex_2007}. Despite such compelling indications for the presence of critical VF in proximity to a CEP, additional signatures of the putative VT are still highly desirable, in order to consolidate the decisive role of the $4f$ electron delocalization in the properties of CeCu$_2$Si$_2$, and accordingly in other Ce-based compounds.

This challenge is strongly constrained by the necessity of both extreme \emph{and} reliable conditions. Here we report for the first time on a \emph{multiprobe} experiment under high pressure, sensing simultaneously electric resistivity $\rho$, \textsc{Hall} $\rho_{xy}$ and \textsc{Nernst} $N$ effect, thermopower $S$ and ac specific heat $C^{ac}$ on the very same CeCu$_2$Si$_2$ single crystal. Based on the resistivity data, we find that the VT is just missed in CeCu$_2$Si$_2$, with a CEP at slightly negative $T$, and suggest a simple method to track the valence cross-over within the $p-T$ plane. More strikingly, it is possible to understand the complete set of low $T$ resistivity data as governed by the existence of the CEP via a simple scaling analysis. In addition, anomalies in the $p$-dependence of $\rho_{xy}$, $S$ and $N$ close to $p_V$ are uncovered and appear intimately related to the Ce valence cross-over (VCO).

\section{Experiment and sample quality\label{}}

\begin{figure}[tb]
\includegraphics[width=\linewidth]{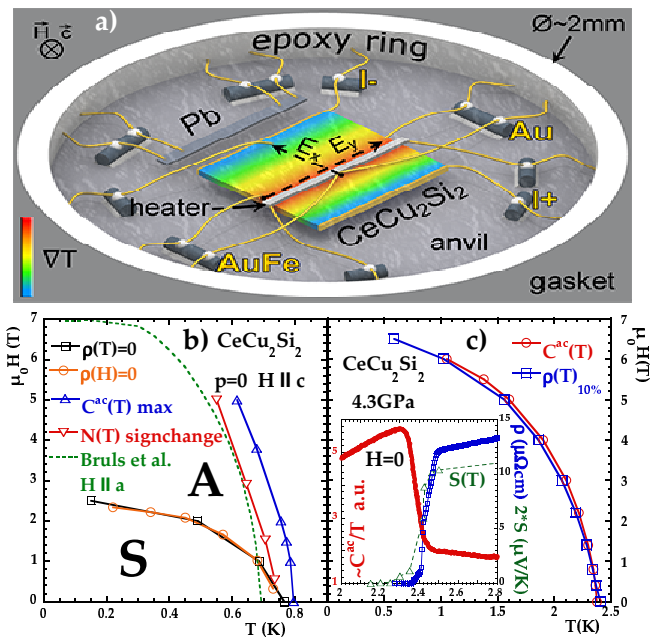}
\caption{a) Schematic view of the pressure cell and all the sensing contacts (Au wires of diameter 13~$\mu$m) on the CeCu$_2$Si$_2$ crystal. b) $H-T$ phase diagram at $p=0$. Superconducting $T_c$ ($\rho$) and $T_N$ ($C^{ac}$) coincide (slightly depending on criteria) for $H=0$, both $C^{ac}$ and $N$ indicate the $A$-phase boundary under field (close to former results from Bruls \textit{et al.}\cite{Bruls_1994}). c) High sample homogeneity at 4.3~GPa, as testified by simultaneous sc transitions at $H=0$ (inset) and under field in $\rho(T)$, $C^{ac}(T)$ and $S(T)$.}
\label{fig2}
\end{figure}

The \emph{multiprobe} configuration (fig.~\ref{fig2}a) offers the unique possibility to analyze complementary properties on the very same sample under \emph{uniform} conditions. Multiprobing is particularly promising for high pressure studies, and in the case of delicate reproducibility. Of course, it is impossible to probe all 5 quantities in an ideal manner: here the geometry has been optimized for $\rho_{xy}$ and $N$ (see ref. \onlinecite{Rueetschi_2011}). Fig. \ref{fig2}a depicts the specifically designed setup within the pyrophyllite gasket of our Bridgman-type tungsten carbide anvil pressure cell \cite{Rueetschi_2007}. Daphne oil 7373 is used as $p$-transmitting medium, providing a high level of hydrostaticity as verified by the included Pb manometer ($p$-gradient of at most 0.1~GPa at 7~GPa). Either an electric current (injected via the whole cross-section of the sample by the wires labeled $I^+$ and $I^-$ in fig.~\ref{fig2}a) or a thermal gradient (created via a resistive heater) can be applied to the ab-plane of the plate-like CeCu$_2$Si$_2$ single crystal ($\sim600\times575\times25\mu \text{m}^3$, same batch as in ref. \onlinecite{Vargoz_1998}). As indicated in fig.~\ref{fig2}a the magnetic field was aligned perpendicular to that plane, along the c-axis. EM DC nanovoltmeters were used to probe the longitudinal and transverse electric fields, $E_x$ and $E_y$ respectively. The $T$-elevation $\Delta T$ with respect to the pressure clamp (placed within a dilution refrigerator) is probed with an Au-AuFe thermocouple (calibrated under magnetic field ($H$) in a separate experiment). For the $C^{ac}$ measurements, the same lock-in based technique as described in ref.~\onlinecite{Alex_thesis} was used for the detection of the amplitude of the periodic thermocouple voltage $V^{ac}$, including separate runs at low frequency to estimate the corresponding DC $T$ elevation. The typical working frequencies varied from about 10~Hz at ambient pressure up to several 100~Hz at high pressures; no sc signal could be detected down to the lowest $T$ from 4.9~GPa onward. For the determination of the thermoelectric power $S=E_x/-\nabla_x T$, we probed the longitudinal field via two voltages, $V_{\text{Au}}$ and $V_{\text{AuFe}}$ (both compared to the opposite longitudinal voltage lead, close to the negative current lead $I^-$ in fig.~\ref{fig2}a). The \textsc{Seebeck} coefficient of the sample can then be obtained from $S=S_{\text{AuFe}}+\tfrac{S_{\text{Au}}-S_{\text{AuFe}}}{1-V_{\text{Au}}/V_{\text{AuFe}}}$, where $S_{\text{Au}}$ can generally be neglected for $T<4.2$~K. This procedure is carried out for several heating currents (different $\Delta T$) and then extrapolated $S(\Delta T\rightarrow 0)$, yielding $S$ for a given cryostat $T$. The \textsc{Nernst} signal $N$ is defined as the transverse electric field generated by a longitudinal $T$ gradient in the presence of a perpendicular magnetic field, $N=E_y/-\nabla_x T\simeq V_y/\Delta T$. The $N$ data were always anti-symmetrized with respect to $H$, and the sign was determined using the common ``vortex convention'', yielding a positive $N$ for moving vortices \cite{Wang_2003,Ong_2006,Rueetschi_2011}. Owing to the experimental configuration we measure here the \textit{adiabatic} \textsc{Nernst} signal with a vanishing transverse heat current, neglecting a possible transverse $T$ gradient \cite{Blatt_1957}.

Sample quality and characterization are particularly important in the case of CeCu$_2$Si$_2$ because of its strong sensitivity on crystal growth conditions and stoichiometry \cite{Steglich_1996}. Here we summarize the key features of our single crystal with the help of 2 figures (at ambient and under $p$) comparing position, shape and width of phase transitions as probed by different physical properties in zero and under magnetic field. Fig.~\ref{fig2}b displays the $H-T$ phase diagram at ambient $p$ obtained from $\rho$, $N$ and $C^{ac}$. The high superconducting $T_c(\rho=~0)=780$~mK goes along with a remarkably low residual $\rho(T_c^+)=8.3~\mu\Omega\text{cm}$ (see inset of fig.~\ref{rho_all}). The sample can be classified as of A/S type \cite{Steglich_1996}, since the vanishing $\rho(T)$ coincides with a pronounced anomaly in $C^{ac}(T)$, corresponding to the antiferromagnetic \textsc{N\'eel} temperature $T_N$ (as identified by its decoupling from $T_c$ under magnetic field and by the associated signchange of $N$, ref. \onlinecite{Rueetschi_2011}). Under pressure, $T_N$ is rapidly suppressed \cite{Lengyel_2009}, and close to $p_V$ (fig.~\ref{fig2}c) a high degree of homogeneity is testified by simultaneous signatures in $\rho(T)$, $S(T)$ and $C^{ac}(T)$ of a \emph{record} superconducting $T_c(H=0)\sim2.4~\text{K}$. These outstanding sample characteristics, in particular the low residual resistivity and the record \emph{bulk} $T_c$, constitute a solid basis for the following data analysis, focussed on the vicinity of $p_V$.

\section{Results and Discussion\label{}}

\begin{figure}[tb]
\includegraphics[width=\linewidth]{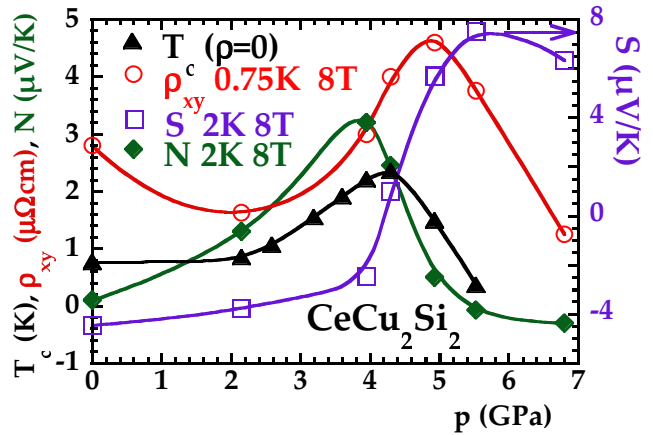}
\caption{$p$-dependence of $T_c^{\rho=0}$, $\rho_{xy}(0.75~\text{K},8~\text{T})$, $S(2~\text{K},8~\text{T})$ and $N(2~\text{K},8~\text{T})$ (lines are guides to the eye). The maximum $T_c$ is accompanied by a pronounced peak in $N(p)$ and a sign-change in $S(p)$, whereas the maximum in $\rho_{xy}(p)$ occurs at slightly higher $p$, more following $\rho_0(p)$ (see fig.~\ref{fig4}b). \label{multi}}
\end{figure}

In the following, we will expose and discuss one by one our results on the different investigated physical properties. As a highlight, we display in fig.~\ref{multi} (which we will often refer to) the $p$-dependence of the superconducting $T_c^{\rho=0}$, and for the first time of $\rho_{xy}(0.75~\text{K},8~\text{T})$, $S(2~\text{K},8~\text{T})$ and $N(2~\text{K},8~\text{T})$. A strong sensitivity of CeCu$_2$Si$_2$ to pressures around 4~GPa~($\sim p_V$) is revealed, as will be developed further throughout this communication. Most emphasis will be put on our electric transport results, since the improved data and sample quality allow us to go beyond previous studies and address quantitatively features linked to the 4f electron delocalization. Subsequently, the other multiprobe results will be presented, partly consolidating already existing data, partly opening unprecedented routes of exploration. Even if at this stage a deep physical analysis may be premature in the latter case, this novel extensive set of data which allows a direct comparison of 5 experimental probes on the very same CeCu$_2$Si$_2$ single crystal under high pressure may be worthwhile on its own, and trigger further experimental and theoretical investigations.

\subsection{Resistivity\label{sec_rho}}
\subsubsection{Comparison to previous results}

\begin{figure}[tb]
\includegraphics[width=\linewidth]{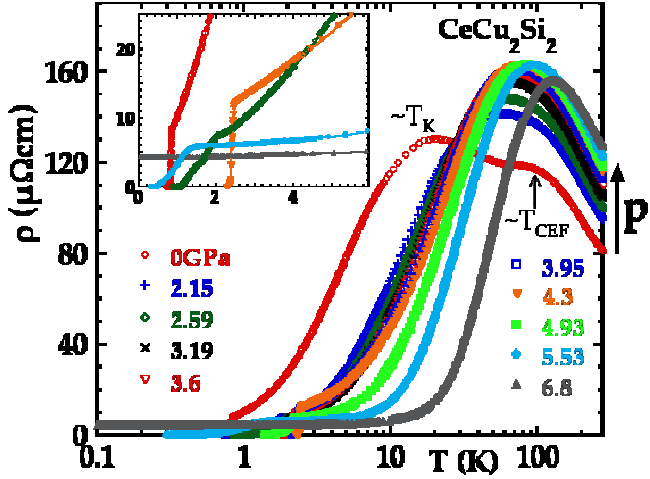}
\caption{Complete $\rho(T,H=0)$ scans for all 10 investigated $p$ on a semi-log scale. At $p=0$, $\rho(T)$ clearly exhibits 2 separated maxima (merging at higher $p$), corresponding to the \textsc{Kondo} coherence ($T_K\sim10$~K) and the crystal electric field ($T_{\text{CEF}}\sim100$~K) energy scales. These data served as a basis to obtain the $p$-dependencies shown in fig.~\ref{fig4}. Inset (same symbols): zoom on the low $T$ region for some selected $p$. Several features become obvious: the increasing superconducting $T_c$ and $\rho_0$ when reaching $\sim4$~GPa, the almost $T$-linear resistivity at 4.3~GPa and the collapse of the $A$ coefficient at highest $p$.}
\label{rho_all}
\end{figure}

 The complete set of electric transport data $\rho(T,p)$ is shown in fig.~\ref{rho_all}. All major characteristics are in line with previous reports \cite{Didier_1999,Alex_PRB_2004,Alex_2007}, namely the extracted $p$-dependence of the parameters $\rho_0$, $A$ and $n$, obtained from simple fits to $\rho(T\leq4.2$~K$, H=0)=\rho_0+A\ast T^n$. In particular, from the inset of fig.~\ref{rho_all} at low $T$, three qualitative features can already be distinguished by eye: as a function of increasing $p$, the extrapolated residual resistivity $\rho_0$ increases before a final downturn, the $A$ coefficient drops dramatically at high $p$ and, maybe less obvious, at 4.3~GPa $n(p)$ has a minimum. The VF theory \cite{Miyake_2007,Alex_2007} predicts for a $4f$ electron delocalization realized at $p=p_V$ a maximum in the associated fluctuations at slightly lower $p$ ($p\lesssim p_V$), coupled to $n\sim1$ and an enhancement in $A$ and~$T_c$. Right at $p=p_V$, the impurity scattering cross-section should rise strongly, leading to a maximum in $\rho_0(p)$. Beyond the delocalization ($p>p_V$), obviously a deep loss of correlation effects is expected. These features are all in good agreement with the experiment (fig.~\ref{fig4}b): while $A(p)$ and $T_c(p)$ culminate around 4~GPa, the extremum in $\rho_0(p)$ points to a slightly higher $p_V$. Note that compared to previous studies, we do not only observe a sudden drop in $A(p)$ (revealing the loss of correlations), but a preceding local maximum which coincides with that in $T_c(p)$ (such a signature of enhanced correlations close to $p_V$ is in line with a significant increase of the initial slope of the upper critical field $|dH_{c2}/dT|_{T=T_c}$ probed by $C^{ac}(p\rightarrow p_V)$, fig.~\ref{fig4}b). Overall, these results corroborate the VF scenario in a general manner, but do not allow to identify the CEP and the VT or VCO line.

\subsubsection{Isothermal $p$-dependence and $4f$ electron delocalization - valence cross-over line and location of the CEP}

\begin{figure}[tb]
\includegraphics[width=\linewidth]{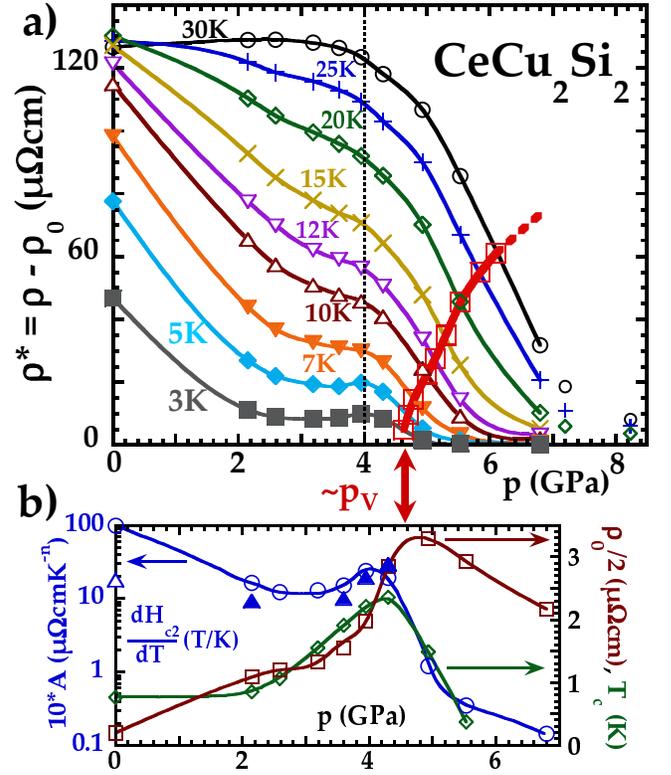}
\caption{a) Isothermal $\rho^{*}(p)=\rho-\rho_0$, constructed from the $\rho(T)$ scans from fig.~\ref{rho_all} at selected $T$ (lines are guides to the eye). The $\rho^{*}(p)$ curves decrease strongly above 4~GPa, within a $p$ range that gets narrower with decreasing $T$. The red open squares indicate the 50\% drop of $\rho^{*}$ compared to the value at 4~GPa (vertical dotted black line). The unconnected data points at high $T$ stem from previous work \cite{Alex_thesis} and solely illustrate how the $\rho^{*}$-downturn completes at high $p$. b) $p$-dependence of the superconducting $T_c$ ($\rho=0$), $\rho_0$ (for visibility scaled by a factor of 0.5), the $A$ coefficient (scaled by a factor of 10) and of the initial slope $|dH_{c2}/dT|_{T=T_c}$ (open ($\rho$) and filled ($C^{ac}$) triangles). For $T\rightarrow 0$ the red VCO line $p_{\text{VCO}}(T)$ from panel a) joins the maximum in $\rho_0(p)$, whereas the maximum of $A(p)$ (and of $|dH_{c2}/dT|_{T=T_c}$) coincides with the maximum $T_c$.}
\label{fig4}
\end{figure}

In order to progress on this issue, we take advantage of the high sample quality, and extend the previous analysis by simply plotting in fig.~\ref{fig4}a the $p$ dependence $\rho^{*}(p)=\rho-\rho_0$ for several $T\leq30$~K (based on our $T$-scans from fig.~\ref{rho_all}). In the following, we want to carefully examine how the $p$-induced $4f$ electron delocalization acts on the resistivity at various $T$. To facilitate the comparison of the different curves, it is essential to subtract the $T$-independent ``background'', i.e. the contribution from impurity scattering $\rho_0(p)$, even if this procedure may introduce some error bars due to the uncertainty on the determination of $\rho_0$ (of less than the symbol size in fig.~\ref{fig4}b). Note that the $p$-dependence of $\rho_0$ is not artificial at all, but can be well understood within the VF theory, as already pointed out above. Furthermore, it is noteworthy that up to $\sim30$~K the phonon contribution remains negligible, as well as the influence of the higher crystal electric field (CEF) levels. However, at higher $T$, the situation may be less straightforward, particularly under $p$ when the ground state degeneracy is affected by the crossing of the \textsc{Kondo} and CEF energy scales (see also fig.~\ref{rho_all}).

\begin{figure}[tb]
\includegraphics[width=\linewidth]{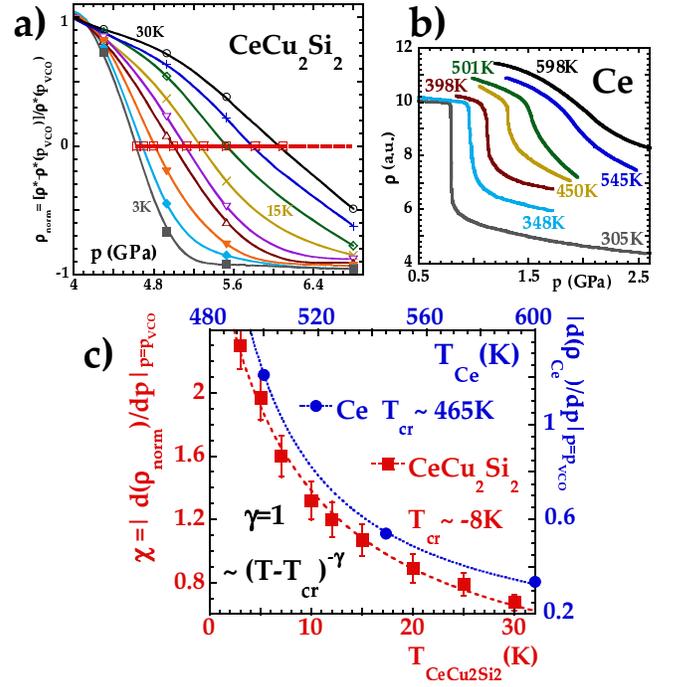}
\caption{a) A simple way for demonstrating that resistivity gets more sensitive to $p$ with lowering $T$ is to normalize by the midpoint of the $\rho^*(p)$-drop (identified by the red open squares in fig.~\ref{fig4}a), compared to the initial value at 4~GPa, yielding $\rho_{\text{norm}}=(\rho^{*}-\rho^{*}(p_{\text{VCO}}))/\rho^{*}(p_{\text{VCO}})$ (lines are guides to the eye). b) Original resistivity data ($p$ scans at fixed $T$) from ref.~\onlinecite{Jayaraman_1965} in elementary Ce. Below the CEP at 480~K, $\rho(p)$ exhibits a clear discontinuity at the first-order valence transition (FOVT), whereas the $\rho(p)$ curves get smoother for $T>T_{cr}$. c) Divergence for $T\rightarrow T_{cr}$ of the slope $\chi=|d\rho_{\text{norm}}/dp|_{p=p_{\text{VCO}}}$, obtained from a) (the error bars take into account the uncertainty on $p$ and that introduced via $\rho_0$). From a simple fit (dashed line), $\gamma\sim1$ and $T_{cr}^{\text{CeCu$_2$Si$_2$}}\sim-8$~K can be extracted. For Ce data \cite{Jayaraman_1965}, a similar treatment yields $T_{cr}^{\text{Ce}}\sim465$~K.}
\label{fig5}
\end{figure}

Let us now look in detail at fig.~\ref{fig4}a: the initial decrease of $\rho^{*}$ with $p$ can be explained with the increase of the coherence (\textsc{Kondo}) energy scale (starting at $\sim 20~\text{K}$ at $p=0$, see fig.~\ref{rho_all}). Preceded by a small upturn at low $T$, due to the enhanced $A$ (fig.~\ref{fig4}b), $\rho^{*}(p)$ drops significantly above $4~\text{GPa}$, indicating the Ce $4f$ electron delocalization. This resistance loss takes place over a $\sim4$ times wider $p$ range at 30~K compared to 3~K. For illustration purposes only we have completed the curves at higher $T$ by data above 7~GPa (isolated points in fig.~\ref{fig4}a), originating from a previous experiment \cite{Alex_thesis} and reanalyzed in the same manner than the present one. Although the overall behavior under $p$ is very similar, we do not include these older data in the further analysis to avoid any ambiguity. The only purpose here is to document the experimentally well-established fact that the delocalization ``terminates'' with a low resistivity, even at 30~K. The observed broadening of $\rho(p)$ with increasing $T$ is reminiscent of elementary Ce \cite{Jayaraman_1965} (see fig.~\ref{fig5}b), where the first-order discontinuity in $\rho(p, T<T_{cr})$ disappears when crossing the CEP ($T_{cr}^{\textnormal{Ce}}\approx 480~\text{K}$ \cite{Lipp_2008}, see inset of fig. \ref{fig1}) and becomes smoother and smoother within the VCO regime. However, in CeCu$_2$Si$_2$, the situation is less straightforward: below 30~K the scattering rate has a strong $T$-dependence. More quantitatively, in Ce the typical magnitude of resistivity on the low $p$ side varies by only $\sim10$\% between 450 and 600~K \cite{Jayaraman_1965} (see fig.~\ref{fig5}b), whereas it varies by more than a factor of 10 in CeCu$_2$Si$_2$, considering the $T$ effect at 3~GPa for example (fig.~\ref{fig4}a). In order to disentangle the intrinsic effect of the delocalization from that of scattering, we consider the \textit{relative} $\rho^{*}(p)$-drop at each $T$, i.e. $\rho_{\text{norm}}=(\rho^{*}-\rho^{*}(p_{\text{VCO}}))/\rho^{*}(p_{\text{VCO}})$ (see fig.~\ref{fig5}a), where $p_{\text{VCO}}$ designates for each $T$ the 50\%-midpoint of the $\rho^{*}(p)$-drop (red squares in figs.~\ref{fig4}a and \ref{fig5}a), compared to the ``initial'' value at 4~GPa (dashed black line in fig.~\ref{fig4}a). Such a normalization can be regarded as a simple way to separate the $T$-dependent contribution to the scattering rate, in order to better detect the $p$-dependence of the \textsc{Drude} weight. Albeit the admittedly small number of data points, the resistance drop is clearly more pronounced at lower $T$ (fig.~\ref{fig5}a), similar to the approach of the CEP in Ce \cite{Jayaraman_1965} (fig.~\ref{fig5}b). Obviously, the chosen criterium of normalization may appear somehow arbitrary, and the critical reader may wonder whether it affects the aforementioned data treatment. Hence we have checked that other criteria like a normalization with respect to 3.5, 4.5 or 5~GPa do not qualitatively alter the main conclusions, i.e. a steeper drop in the normalized resistivity with lowering $T$. The reference point at 4~GPa, just at the onset of the downturn, simply appears to be the most intuitive and reasonable one.

In any case, resistivity gets more sensitive to $p$ for decreasing $T$, consistent with the approach of a CEP, at which the resistance drop should get vertical, like in elementary Ce. In order to locate the CEP, we quantified the steepness of the resistance drop through its slope at the midpoint (red squares in figs.~\ref{fig4}a and \ref{fig5}a), $\chi=|d\rho_{\text{norm}}/dp|_{p=p_{\text{VCO}}}$, and analyzed its divergence $\chi(T\rightarrow T_{cr})$, as shown in fig. \ref{fig5}c. The error bars take into account the uncertainty on $p$ and that introduced via $\rho_0$ (mentioned above). A simple fit of the type $\propto (T-T_{cr})^{-\gamma}$ yields $T_{cr}^{\textnormal{CeCu$_2$Si$_2$}}\cong -8\pm3~\text{K}$ and $\gamma\cong1$, and a similar treatment of the Ce data \cite{Jayaraman_1965} $T_{cr}^{\textnormal{Ce}}\cong 465~\text{K}$ (in agreement with more recent data \cite{Lipp_2008}). Notice that experimental limitations like sample imperfections or the inevitable $p$-gradient tend to broaden the resistivity drop, and hence the extracted $T_{cr}$ may be slightly underestimated. In this respect, locally a FOVT (slightly positive $T_{cr}$) may not be excluded, so that density fluctuations may mediate SC \cite{Monthoux_2004} via a vanishing compressibility (ref.~\onlinecite{Lipp_2008}, Ce case).

This basic analysis of the slope $\chi(T\rightarrow T_{cr})$ establishes for the first time experimentally that the CEP of the VT line in CeCu$_2$Si$_2$ lies at a slightly negative $T$. Even if some uncertainty on the precise value remains - which is almost unavoidable in such kind of high $p$ experiment - this outcome is more than just a qualitative result, in the sense that it clearly specifies the (so far unknown) magnitude of $T_{cr}^{\textnormal{CeCu$_2$Si$_2$}}$ to be of the order of several degrees. It substantiates that the FOVT is just missed in CeCu$_2$Si$_2$, meaning that only a VCO is realized. Of course, electric transport on its own constitutes only an indirect measure of the valence change. But recent reports \cite{Rueff_2011, NQR_2008} based on more microscopic probes of the $4f$ electron count, and their qualitative conclusions on a possible VCO in CeCu$_2$Si$_2$ under $p$, corroborate our results and assumptions. Furthermore, previous $\rho$ data on the CeCu$_2$(Si$_{1-x}$/Ge$_x$)$_2$ system (shown in fig.~4 of ref.~\onlinecite{Yuan_2006}) exhibit for $x=0.1$ qualitatively similar features than those presented here. Altogether, a consistent scenario emerges, in which the key ingredient is the proximity of CeCu$_2$Si$_2$ at $p_V$ to the CEP: this constellation allows the associated critical VF to still play an important role in the low $T$ physics, such as mediating SC, in very close agreement to the already introduced VF theory (see in particular the scenario outlined in fig. 2b of ref. \onlinecite{Wata_2011}). In this respect, our results do not only specify the location of the CEP close to $T=0$, but offer also a simple method to track the VCO within the $p-T$ plane: connecting the previously defined midpoints (red squares in figs.~\ref{fig4}a and \ref{fig5}a), we obtain the red VCO line drawn in fig.~\ref{fig1} (equivalent to that in fig.~\ref{fig4}a). As already mentioned, the precise pathway may slightly depend on the normalization criteria. On this footing, the minor downward curvature on the high $T$ side of the VCO line may be somehow exaggerated: at 4~GPa and 30~K, resistivity obviously saturates, approaching more or less the unitary limit. This means that the midpoint should actually be situated at a marginally lower $p_{\text{VCO}}$. Regardless of this detail, and compared to the rather qualitative lines in previous $p-T$ phase diagrams \cite{Yuan_2003,Didier_2005,Yuan_2006,Monthoux_2007}, the present one is the first signature of the $4f$ electron delocalization based on experiment. Finally, extrapolating this line, $p_{cr}^{\textnormal{CeCu$_2$Si$_2$}}(=p_V)\cong4.5\pm0.2~\text{GPa}$ can be deduced, so that both coordinates of the CEP are determined within the experimental errors. Joining figs.~\ref{fig4}a and b, it becomes obvious that the extrapolation of the red VCO line roughly coincides with the maximum in $\rho_0(p)$, which is a good sign of consistency (within the framework of the critical VF), since both features are triggered precisely by the $4f$ electron delocalization.

\begin{figure}[tb]
\includegraphics[width=\linewidth]{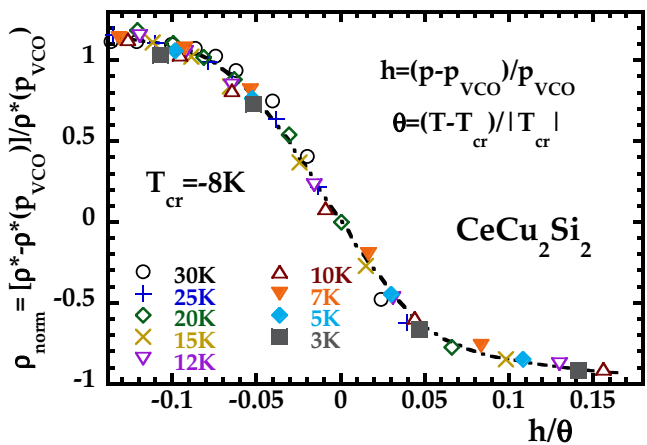}
\caption{Best collapse of all normalized $\rho_{\text{norm}}(p)$ data (from fig.~\ref{fig5}a) as a function of the generalized ``distance'' $h/\theta$ from the CEP ($T_{cr}\sim-8$~K, $p\geq3.6$~GPa, dashed line: guide to the eye). Within the universal framework of critical phenomena (see text), such a representation corresponds to $\gamma=1$ and $\delta\rightarrow\infty$.}
\label{fig6}
\end{figure}

\subsubsection{Uniform behavior at a generalized distance from the CEP}

Next let us focus on an additional feature, revealing in a complementary manner that resistivity is governed by the existence of an underlying CEP. Strikingly, it is possible to collapse all the $\rho_{\text{norm}}(p)$ data on a single curve in vicinity to $p_{\text{VCO}}$. Introducing the dimensionless variables $h=(p-p_{\text{VCO}})/p_{\text{VCO}}$ and $\theta=(T-T_{cr})/|T_{cr}|$, a collapse of excellent quality is obtained according to $\rho_{\text{norm}}=f(h/\theta)$ ($f$: scaling function) with $T_{cr}\cong -8\pm5~\text{K}$, as shown in fig.~\ref{fig6} for $p\geq3.6$~GPa. This means that for a generalized ``distance'' $h/\theta$ from the CEP, the $\rho_{\text{norm}}$ data behave in a unique manner. As indicated, the sensitivity on the chosen $T_{cr}$ is limited as well as the number of data points, and the applicable $p$ range may slightly vary according to the normalization reference (see above). Nevertheless, the data collapse is quite robust and its quality noteworthy, as well as the fact that the extracted $T_{cr}$ is of the same order of magnitude as previously deduced from the $T$-divergence of $\chi$. We have also tried to apply the universal scaling theory of critical phenomena, which has for example been successful in describing the \textsc{Mott} transition probed by $\rho(p,T)$ in oxide materials \cite{Limelette_2003} or in characterizing the valence instability in Ce$_{1-x}$Th$_x$ \cite{Lawrence_1975}. Within such a generalized framework, the expected scaling law for $T>T_{cr}$ yields: $\rho_{\text{norm}}/h^{1/\delta}=f(h/\theta^{\gamma\delta /(\delta -1)})$  with $\gamma$, $\delta$: critical exponents (mean field approach: $\gamma^{\text{MF}}=1$, $\delta^{\text{MF}}=3$). In the case of CeCu$_2$Si$_2$ this leads empirically to $\gamma=1$ and $\delta\rightarrow\infty$ (fig.~\ref{fig6}). Even if the collapse is not strongly sensitive on the choice of $\delta$, it seems incompatible with $\delta<5$ and does hence not correspond to a simple universality class such as the liquid-gas transition. Further experiments with a higher data density, also in related compounds, may help to better understand this issue.

\subsubsection{General relevancy of $\rho(p)$ analysis}

Altogether, the divergence of the slope $\chi(T\rightarrow T_{cr})$ and the observed data collapse strongly suggest that the resistivity around $p_V$ is governed by the proximity to a CEP (at slightly negative $T$). This outcome offers different avenues for future exploration. First, it may be interesting to study on CeCu$_2$Si$_2$ single crystals of different origin the involvement of critical VF in SC, i.e. the relationship between $T_c$ and $T_{cr}$ (of course other factors that influence $T_c$ have to be taken into account as well). Second, even more promising, the widespread relevance of the VF scenario should be investigated by applying a similar resistivity analysis to other Ce-based compounds, especially to the more common cases where $p_c$ and $p_V$ are not well separated. Indeed, the identification of $p_V$ and the $4f$ electron delocalization is generally more elusive than that of the breakdown of magnetism, $p_c$, reflecting the competition between the \textsc{Kondo} effect and the RKKY interaction. Rather exceptionally, in CeCu$_2$Si$_2$ $p_c\ll p_V$, and $\rho(T,p=0)$ exhibits two distinct maxima/energy scales, as already noticed in fig.~\ref{rho_all} ($T_K$ and $T_{\text{CEF}}$). Strikingly, close to $p\rightarrow p_V$ these maxima merge. Although the microscopic link with the VCO may not be clear-cut, phenomenologically such merging maxima in resistivity can be found in numerous other systems (like CePd$_2$Si$_2$/Ge$_2$ \cite{Wilhelm_2002}, CeRu$_2$Ge$_2$ \cite{Wilhelm_2004} and CeCu$_5$Au \cite{Wilhelm_1999}), where the proximity to an underlying CEP may therefore be revealed. However, in contrast to CeCu$_2$Si$_2$, in those systems both maxima already overlap for $p\rightarrow p_c$. Hence $p_c\lesssim p_V$ is suspected. The latter scenario is also assumed in CeRhIn$_5$, mainly based on the subsistence of SC in a wide region around $p_c$, the maximum in $\rho_0$ and a $T$-linear resistivity ($n\sim1$) \cite{Muramatsu_2001,Knebel_2008,Park_2008}. Using our method could be an important ingredient to probe the existence of a CEP and the relevancy of VF, notably for SC, in these compounds. Nonetheless, another dissimilarity to CeCu$_2$Si$_2$ may occur: the resistance drop associated with the delocalization may be less pronounced, which means experimentally less accessible, due to a lower lying CEP (more negative $T_{cr}$). Concurrently, the impact of the critical VF on the physical properties gets reduced. Previous experimental data in other Ce systems (obtained mainly under less reliable conditions) may in some cases (for example in CeCu$_2$Ge$_2$ \cite{Vargoz_1998_JMMM,Vargoz_thesis}) exhibit close to $p_V$ features qualitatively similar to those discussed here in CeCu$_2$Si$_2$, but do not allow, so far, an incontrovertible quantitative analysis. New high $p$ experiments with improved accuracy will facilitate progress on this explorative route. Let us finally add that in other systems like CeIrIn$_5$ \cite{Capan_2004,Capan_2009} or YbAg/AuCu$_4$ \cite{Sarrao_1999,Wada_2008} $p$ is replaced by magnetic field as tuning parameter to induce the proximity to a CEP and critical VF.

\subsubsection{Broadening of the resistive transitions}

Last but not least, let us comment on the broadening of the sc transitions as seen by $\rho(T)$ at intermediate and high $p$. The inset of fig.~\ref{rho_all} exemplifies typical shapes of dropping resistivities at 2.59 and 5.53~GPa. In the $p-T$ phase diagram (fig.~\ref{fig1}), we have represented two lines delimiting the superconducting phase, one corresponding to the onset temperature ($T_c^{\text{onset}}$) and the other to $\rho=0$ ($T_c^{\rho=0}$). The resulting separation of both lines \textit{below} and \textit{above} $p_V$ (in conjunction with narrow transitions at $p=0$ and $p_V$) is robust and found systematically in all CeCu$_2$Si$_2$ (and CeCu$_2$Ge$_2$) samples (already mentioned in early reports \cite{Didier_1985}), and does not depend on the $p$ conditions, since it has been observed in a variety of pressure media\cite{Didier_1999,Alex_PRB_2004,Alex_2007,Didier_2005,Vargoz_1998}, among which in highly hydrostatic $^4$He. One may argue that broad resistive transitions are quite common at the edges of superconducting regions, but in CeCu$_2$Si$_2$, at intermediate $p$, the broadening appears precisely in a region where the $p$-dependence of $T_c$ is rather weak. For fig.~\ref{fig1}, between 0 and 2.15~GPa, older data \cite{Vargoz_1998,Alex_thesis} served as a basis for a smooth interpolation. $T_c^{\text{onset}}(p)$ has already reached the level of the maximum $T_c$ at $\sim2.5$~GPa, but without a cusp-like feature as in previous reports \cite{Thomas_1996,Alex_PRB_2004}. This rather quick increase of $T_c^{\text{onset}}(p)$ may point to the fact that the VF pairing mechanism needs to be considered already at low $p$ (at least in some form), as already inferred from the very weak decrease of $T_c(p)$ above $p_c$ \cite{Lengyel_2011}. Around $p_V$, the maximum $T_c^{\text{onset}}$ and $T_c^{\rho=0}$ coincide. Above $p_V$, in several samples\cite{Thomas_1996,Vargoz_1998,Alex_2005} the superconducting region extends to high $p$, i.e. the decrease of $T_c^{\text{onset}}(p>p_V)$ is less steep than shown in fig.~\ref{fig1}, but may exhibit a jump close to 8~GPa \cite{Thomas_1996}. As a general feature, the bulk $T_c$ as probed by $C^{ac}$ coincides with $T_c^{\rho=0}$. More precisely, at intermediate $p$ $C^{ac}(T)$ does neither display any signature of $T_c^{\text{onset}}$ nor any notable broadening of the superconducting anomaly associated with $T_c^{\text{bulk}}\sim T_c^{\rho=0}$ (not shown, but similar to reports in refs.~\onlinecite{Alex_thesis,Alex_PRB_2004}). Remarkably, above $\sim4.9$~GPa, no more anomaly in $C^{ac}(T)$ can be detected, down to the lowest reachable $T$. In this context, it may be worthwhile to notice that the $T_c$'s from ref. \onlinecite{Thomas_1996} originate from ac susceptibility, sensitive to a surface layer. Altogether, the broadening of the resistive transition at intermediate and high $p$ in CeCu$_2$Si$_2$ appears as an \textit{intrinsic} phenomenon, even if details may be sample dependent. However, the physical origin of this broadening remains an open question. Further investigations are needed to elucidate whether it may be related to the symmetry change of the sc order parameter (at least at intermediate $p$) \cite{Lengyel_2009}, to phenomena observed in the Ce-115 series \cite{Park_2012} or more directly to the presence of critical VF. Indeed, one could imagine a non-homogeneous nucleation and spread of superconducting domains within the sample or on the surface, depending on the local distribution of $T_{cr}$ and the closeness to $p_V$.

\subsection{Magnetoresistance}

\begin{figure}[h]
\includegraphics[width=\linewidth]{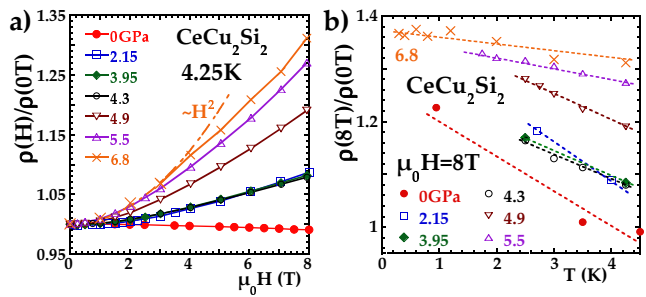}
\caption{a) $H$-dependence of the ratio $\rho(H)/\rho(H=0)$ for various $p$ at 4.25~K. At ambient $p$, in more usual terms the magnetoresistance $\Delta\rho$($=\rho(H)-\rho(H=0)$) is slightly negative, becomes positive for $p>0$, but without any significant evolution between 2 and 4~GPa, and finally increases further. Just for comparison, a curve $\propto H^2$ is also shown as dash-dotted line. b) $T$-dependence of the ratio $\rho(8~\text{T})/\rho(H=0)$ for various $p$. The points at the right correspond to those of panel a) at 8~T. Surprisingly, the extrapolation for $T\rightarrow0$ does not seem to be strongly dependent on $p$. (lines are guides to the eye only)}
\label{MR}
\end{figure}

Let us now study the influence of magnetic field on resistivity, i.e. the magnetoresistance. First we stress that solely $H$-symmetrized $\rho(H)$ data are used in the following discussion, excluding any spurious contribution from \textsc{Hall} effect, but strongly limiting the number of available data points. Fig.~\ref{MR} displays the measured $H$- and $T$-dependencies of the ratio $\rho(H)/\rho(H=0)$ for several $p$, at 4.25~K and 8~T respectively. We preferred the representation of the relative value $\rho(H)/\rho(0)$ to the absolute value $\Delta\rho=\rho(H)-\rho(H=0)$ (commonly called magnetoresistance) because of the $p$-dependence of the residual $\rho_0$. At ambient $p$ and for decreasing $T$ $\Delta\rho$ changes sign from negative to positive at around 4~K, as already reported earlier \cite{Rauchschwalbe_1987} (at about 2~K in ref.~\onlinecite{Aliev_1984}). In a simplified picture, the sign-change may be considered as a rough measure of the coherence temperature $T_{coh}$ of the Ce \textsc{Kondo} lattice (generally $T_{coh}\ll T_K$), as also indicated by the maximum in the \textsc{Hall} data in the same $T$ region (see sect.~\ref{ch_Hall}). However, quantitatively we find a more pronounced $\rho(H)/\rho(0)$ than measured back in 1987 (ref. \onlinecite{Rauchschwalbe_1987}): for example at 1.5~K a linear interpolation  yields $\rho(8~T)/\rho(0)\sim1.15$ (fig.~\ref{MR}b), which corresponds to an increase of at least a factor of 5, despite our very low $\rho_0$. Comparing crystals of various quality (investigated by our group over the last 15 years or so\cite{Vargoz_1998,Alex_thesis}), the absolute magnetoresistance $\Delta\rho$ depends very little on $\rho_0$ at temperatures of a few K and a magnetic field of about 8~T. It looks like as if resistivity had two contributions: one that is field independent and varies strongly from sample to sample, and another one which always gives the same order of magnitude of $\Delta\rho$ in magnetic field. This explains, at least phenomenologically, a considerable spread in $\rho(H)/\rho(0)$ on different samples. Next let us focus on the effect of $p$. At 4.25K (fig.~\ref{MR}), $\Delta\rho(p)$ first gets positive, in accordance with the increase of $T_K$ (and hence of $T_{coh}$) with $p$. Surprisingly, from $\sim2$~GPa onward $[\rho(H)/\rho(0)](p)$ is almost constant, and continues to rise only above $p_V$, before likely reaching some saturation at high $p$. Roughly, such a behavior is reminiscent of $A(p)$ which drops with $p$, except in the region before $p_V$ (where it even gets enhanced, see fig.~\ref{fig4}b). Conceivably, the effect of a raising $T_K(p)$ (resulting in an increasing  $[\rho(H)/\rho(0)](p)$) is superimposed by another phenomenon, of opposite effect and driven by critical VF below $p_V$. Further investigations are necessary to elucidate this question. Considering the $p$-evolution of $\rho($8~T$)/\rho(0)(T)$, we find another unexpected result: the extrapolation for $T\rightarrow0$ seems to point to an almost $p$-independent $\rho($8~T$)/\rho(0)$ of about 1.35: the higher values of $\rho(H)/\rho(0)$ at finite $T$ are ``compensated'' by a weaker $T$-dependence for increasing $p$. Of course, more data points at low $T$ would be needed, especially around $p_V$, to confirm this trend. Less intriguingly, $\rho(H)/\rho(0)$ clearly follows the usual $H^2$-dependence at low fields, as exemplified in fig.~\ref{MR}a for 6.8~GPa. Another standard examination is provided by the validity check of \textsc{Kohler}'s rule. It states that the effect of magnetic field on resistivity, $\Delta\rho(H)/\rho(0)$, scales as a function of $H/\rho(0)$ \cite{Pippard_1989}. Deviations from this standard behavior are also observed, for example in CeRhIn$_5$ where magnetoresistance violates \textsc{Kohler}'s rule in the $p$ region exhibiting prominent non-\textsc{Fermi} liquid properties, but satisfies a modified scaling relation \cite{Nakajima_2007}. From our partial analysis (not shown) on CeCu$_2$Si$_2$, no significant deviation from \textsc{Kohler}'s rule throughout the investigated $p$ range can be deduced. However, this preliminary statement still requires substantiation from supplementary data, especially at high $p$ where $\rho(T<4$~K$)$ hardly varies, meaning that data points above 4~K should be included in order to perform the test on a broad parameter range.

\subsection{Specific heat (ac)}

\begin{figure}[h]
\includegraphics[width=\linewidth]{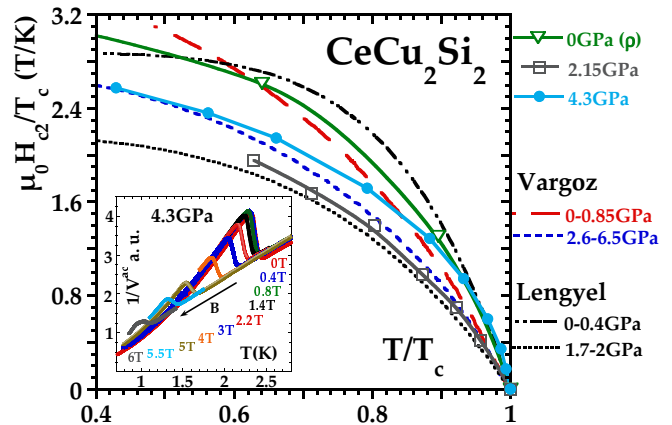}
\caption{Normalized $H_{c2}(T)$ data from $C^{ac}$ (except $p=0$: from $\rho$). Distinct behaviors for $p=0$ and $\sim 2$~GPa are confirmed (compared to a specific heat study by Lengyel \textit{et al.} \cite{Lengyel_2009}), but around $p_V$ another regime appears, with an initial slope $|dH_{c2}/dT|_{T=T_c}$ closer to that at $p=0$ (see also fig.~\ref{fig4}b). Previous resistivity data from Vargoz \textit{et al.}\cite{Vargoz_1998}, available for $p=0$ and $p_V$ for a sample of the same batch, confirm this latest trend. Inset: example of raw data (inverse of the thermocouple voltage) under field at 4.3~GPa, used to determine the different $H-T$ phase diagrams. Note that for $p\geq4.9~$GPa no superconducting anomaly is detected by $C^{ac}$ down to 0.1~K.}
\label{fig7}
\end{figure}

Concerning ac specific heat, its main contribution, apart from validating bulk SC, is to confirm for the first time the enhancement of correlation effects near the maximum $T_c$ by a bulk probe, at least when considering that the initial slope $|dH_{c2}/dT|_{T=T_c}$ qualitatively reflects the quasiparticle velocity or effective mass as a correlation strength sensor. Indeed, an enhancement by a factor of $\sim3$ of the initial slope at the maximum $T_c$ compared to 2~GPa is observed (filled triangles in fig.~\ref{fig4}b). As mentioned earlier, such a trend had already been inferred from resistivity and a kind of ``plateau'' in $A(p\sim4~\text{GPa})$ \cite{Didier_1999,Alex_PRB_2004,Alex_2007}, which has now grown into a clear maximum in the present high quality single crystal, fig.~\ref{fig4}b. More specifically, one may also be interested in the $p$-evolution of the properties of the superconducting phase itself, which is in part characterized by the $H_{c2}(T)$ curves. Without pushing the analysis too far, we just present our new $H_{c2}(T)$ data that allow an extension up to $p_V$ of a scaling study suggested by Lengyel \textit{et al.}\cite{Lengyel_2009} (restricted to $p<2.1$~GPa). Fig.~\ref{fig7} depicts the corresponding normalized $H/T_c-T/T_c$ phase diagram, including the data from ref. \onlinecite{Lengyel_2009}. For $p\sim0$ ($\rho$ data) and $p\sim2~\text{GPa}$ the trend of 2 distinct regimes is confirmed, i.e. a ratio of the initial slopes $|dH_{c2}/dT|_{T=T_c}$ of roughly 1.75 (ref. \onlinecite{Lengyel_2009}). However, for $p\rightarrow p_V$, the initial slope rises again (to even higher values than at $p=0$, see above). It hence appears questionable whether the behavior at only $\sim2$~GPa is representative of the critical VF regime, as implied by the presumed symmetry change of the superconducting order parameter \cite{Lengyel_2009}. On the present data basis, a further evolution with increasing $p$ towards $p_V$ can not be excluded. For sure, the robustness against $H$ is first lowered by $p$, and then rises again for $2~\text{GPa}<p\lesssim p_V$. This trend is supported by an anterior report from Vargoz \textit{et al.}\cite{Vargoz_1998} on a crystal of the same batch (original $\rho$ data represented by red and dark blue lines in fig.~\ref{fig7}), where the initial slopes for $p\sim0$ and $p\sim p_V$ roughly match. Notice that the Pauli limiting field ($T\rightarrow0$) seems slightly weakened for $p\rightarrow p_V$, pointing to an increasing $g$-factor \cite{Vargoz_1998}. Finally, the evolution with $p$ of the superconducting order parameter symmetry may be easier tracked by angle-dependent measurements.

\subsection{Transverse resistivity - \textsc{Hall} effect}\label{ch_Hall}

\begin{figure}[h]
\includegraphics[width=\linewidth]{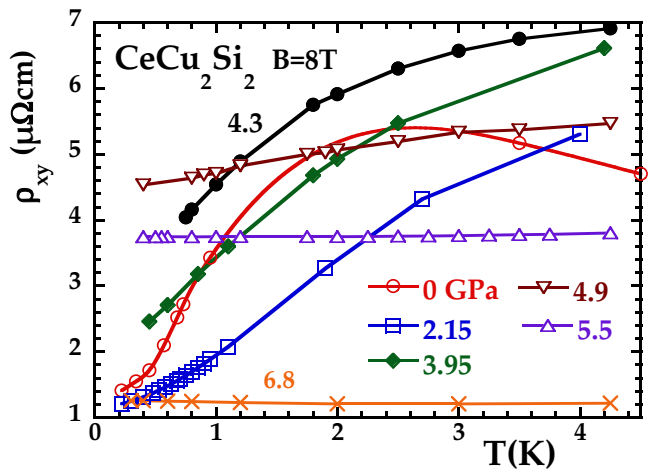}
\caption{$T$-dependence of the transverse resistivity (\textsc{Hall} signal) $\rho_{xy}(8~\text{T})$ at several $p$ up to 6.8~GPa. At $p=0$ a maximum is observed around 2~K, possibly reflecting the $T_{coh}$ energy scale. With increasing $p$ and at low $T$, the positive curvature evolves subsequently into a $T$-linear and a $T$-independent behavior, whereas $\rho_{xy}(T\rightarrow0)$ appears maximum at a $p$ slightly above $p_V$ (displayed in fig.~\ref{multi}), similar to $\rho_0(p)$ (fig.~\ref{fig4}b). (lines are guides to the eye only)}
\label{rxy_T}
\end{figure}

In fig.~\ref{rxy_T} we display the $T$-dependence of the transverse resistivity (\textsc{Hall} effect) $\rho_{xy}(T)$ at 8T and for several $p$ up to 6.8~GPa (data systematically anti-symmetrized with respect to $H$), the $H$-dependence $\rho_{xy}(H)$ is exemplified at $\sim4$~GPa in fig.~\ref{4timesB}a for different $T$. At $p=0$ the measured $T$-dependence is roughly in line with the literature \cite{Aliev_1984,Onuki_1989,Araki_2011} (except for the sign compared to ref.~\onlinecite{Aliev_1984}) with a maximum at $\sim2$~K, possibly pointing to the $T_{coh}$ energy scale, comparable to that inferred from magnetoresistance data (see above). For a precise analysis of the ordinary and the anomalous contribution due to skew scattering, data above 4~K would be necessary and should be compared to magnetic susceptibility measurements \cite{Fert_1987,Araki_2011}. This is particularly true for higher $p$ where we did not reach the maximum in $\rho_{xy}(T)$, shifted to $T>4$~K (ref.~\onlinecite{Araki_2011}) and following the increase of $T_K$ (and hence of $T_{coh}$) with $p$ (see the $p$ shift of the maximum in $\rho(T)$, fig.~\ref{rho_all}). Here we report for the first time on a CeCu$_2$Si$_2$ single crystal the $p$-dependence of $\rho_{xy}$ at low $T$ and across the region of the 4f electron delocalization, where it exhibits a prominent maximum (fig.~\ref{multi}). Its position does not depend on the extraction from high or low field data, due to an almost $H$-linear \textsc{Hall} voltage (fig.~\ref{4timesB}a); its value is about one order of magnitude higher than that of LaCu$_2$Si$_2$ (ref. \onlinecite{Aliev_1983}). As can easily be deduced from fig.~\ref{rxy_T}, the strongest \textsc{Hall} signal corresponds to 4.3 and 4.9~GPa, depending on the $T$ range; below 1~K the maximum in $\rho_{xy}(p)$ clearly lies beyond that of $T_c(p)$, as shown in fig.~\ref{multi} at 0.75~K and 8~T. This shift points to a \textsc{Hall} signal more sensitive to the valence cross-over itself, mimicking $\rho_0(p)$ \cite{Miyake_2007}. In contrast, the \emph{maxima} in $T_c(p)$ and $A(p)$ rather appear for $p\lesssim p_V$, the domain of strongest VF \cite{Miyake_2007} (see above). For $p>p_V$, the decrease of $\rho_{xy}(p)$ is quite steep, like that of $T_c(p)$ or $A(p)$, reflecting the departure from the critical region. The $T$-dependence $\rho_{xy}(T\rightarrow 0)$ also clearly changes over the whole $p$ range (from positive curvature over $T$-linear at $p_V$ towards almost $T$-independent, fig.~\ref{rxy_T}). Interestingly, an enhancement of $\rho_{xy}$ at low $T$ is also reported in CeRhIn$_5$ close to the critical pressure $p_c\sim p_V$, and discussed in the light of the backflow effect in strongly correlated materials \cite{Nakajima_2007}. Further calculations within the critical VF scenario are currently carried out and may connect on a microscopic basis the observed increase of $\rho_{xy}(p)$ in vicinity of $p_V$ with the 4f electron delocalization and the proximity to the CEP in CeCu$_2$Si$_2$.

\subsection{Thermoelectric properties I - thermopower}

\begin{figure}[h]
\includegraphics[width=\linewidth]{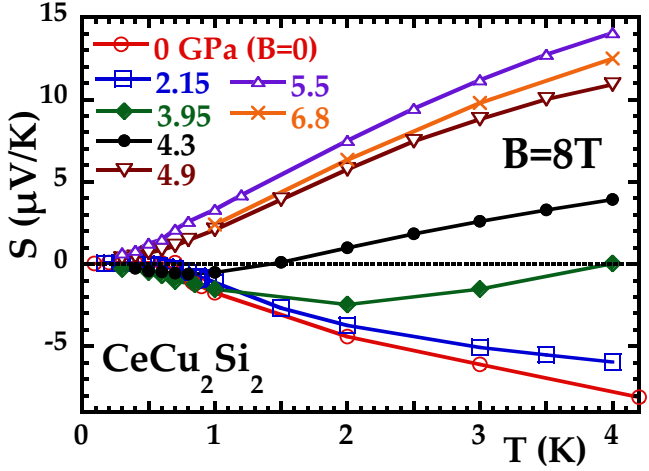}
\caption{$T$-dependence of the thermoelectric power $S(8~$T$)$ at several $p$ up to 6.8~GPa. Remarkably, a sign change from negative to positive takes place around 4~GPa within the entire investigated $T$ range. For $0<p<5\text{GPa}$, $S(8~$T$)$ remains negative down to the lowest available $T$. The $p$-dependent sensitivity on magnetic field is displayed in figs.~\ref{6times}a-c. (lines are guides to the eye only)
\label{TEP_T}}
\end{figure}

Let us now concentrate on the thermoelectric properties, $S$ and $N$. Starting with the (longitudinal) thermopower (\textsc{Seebeck} coefficient), our data (figs.~\ref{multi}, \ref{TEP_T} and \ref{6times}a-c) generally agree with previous reports \cite{Aliev_1984,Sparn_1985,Didier_1985}, except that at $p=0$ we obtain a slightly smaller signal from the $A$-phase (not shown) compared to ref.~\onlinecite{Sparn_1985}. Apart from this small positive contribution below 1~K, the thermopower signal is negative, before getting positive around 70~K with a subsequent maximum at about 150~K, ref.~\onlinecite{Didier_1985}. Even if a generic behavior is absent in the thermoelectric response of HF materials, it seems to be more or less characteristic \cite{Fischer_1989,Link_1996,Zlatic_2003} for Ce compounds close to the magnetic instability that at low $T$ a negative contribution competes with the more ordinary positive contribution (related to the $T_K$ and $T_{\text{CEF}}$ energy scales) to $S$. Hence a reduction of this negative component is expected when $p$ is applied, since an increasing hybridization generally favors the non-magnetic HF \textsc{Kondo} lattice state (screened 4$f$ electrons), eventually followed by a mixed valence state. This phenomenon has already been observed experimentally \cite{Didier_1985} on a broad $T$ scale up to 300~K and 8~GPa, demonstrating the continuous decrease and subsequent vanishing of the negative component. Here we concentrate on the low $T$ region, for the first time in the presence of a magnetic field (to avoid the superconducting shunt). As a key feature, $S(p,2$~K$,8$~T$)$ changes sign precisely in the region where $T_c(p)$ and $A(p)$ are maximum (see figs.~\ref{multi} and \ref{fig4}b): like in a naive picture, $S(p)$ seems related to the derivative of the density of states. Whether this sign-change of thermopower is intimately coupled to the 4f electron delocalization, or occurs rather accidentally in this $p$ region, is left for further studies. In particular, future band structure calculations or \textsc{Fermi} surface investigations under $p$ may help to better identify the role of the VCO for the intriguing behavior of $S(p,T,H)$. Close to 6~GPa, $S(p,2$~K$,8$~T$)$ finally has a maximum itself (fig.~\ref{multi}), the down-turn occurring only at the highest examined $p$. Surprisingly, our results seem not compatible with previous systematics \cite{Kamran_2004}, which would expect $S/T(T\rightarrow 0)\sim N_Ae/\gamma$, where $N_Ae=9.6\times10^5~\text{C~mol}^{-1}$ is the \textsc{Faraday} number and $\gamma$ the \textsc{Sommerfeld} coefficient of specific heat. From our data in the vicinity of $p_V$, $S$ remains negative down to the lowest available $T$ (for $H\neq0$), except for $p>5$~GPa. Possibly, positive values are recovered at much lower $T$, since the characteristic energy scale of the system may be markedly reduced around $p_V$. On the same footing, the \textsc{Seebeck} coefficient displays a strong $H$-dependence in the vicinity of $p_V$, in contrast to lower and higher $p$, as shown in figs.~\ref{6times}a-c. The difference in the sensitivity on magnetic field is particularly striking between 4~GPa with $|S(8~T)-S(0~T)|/S(0~T)\sim 1.75$ at 3~K (fig.~\ref{6times}b) and the $H$-independent $S(T)$ at 6.8~GPa (fig.~\ref{6times}c). This is to our knowledge the first observation of such a feature in $S(H)$ close to a valence instability, and it clearly points to a low characteristic energy scale and a peculiar behavior around $p_V$, as already mentioned above; however, its physical origin may be difficult to unveil independently from other probes since it involves both transport and thermodynamic properties of the system.

\subsection{Thermoelectric properties II - \textsc{Nernst} effect}

\begin{figure}[h]
\includegraphics[width=\linewidth]{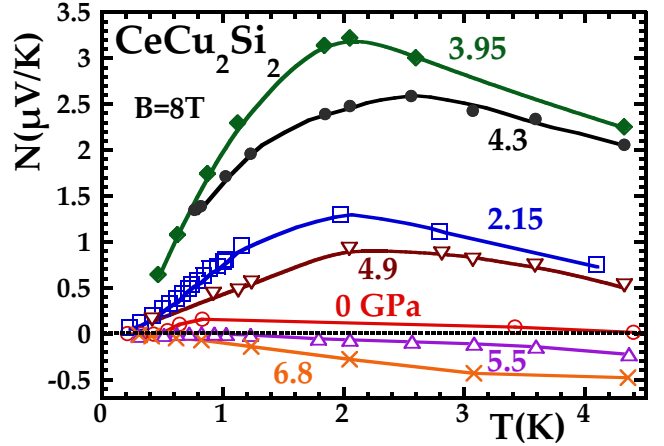}
\caption{$T$-dependence of the \textsc{Nernst} signal $N(8~$T$)$ at several $p$ up to 6.8~GPa. As a function of $p$, the \textsc{Nernst} signal first increases, with a maximum at 4~GPa (see fig.~\ref{multi}), before decreasing again and even switching to negative values at high $p$. Below 5~GPa, a characteristic energy scale may be revealed through the maximum in $N(T)$ around 2~K. (lines are guides to the eye only)}
\label{N_T}
\end{figure}

\begin{figure}[h]
\includegraphics[width=\linewidth]{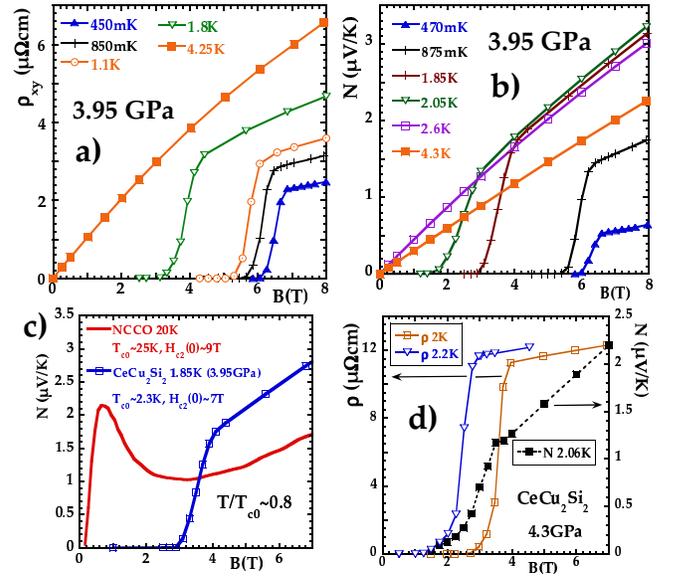}
\caption{a) $H$-dependence of the \textsc{Hall} signal $\rho_{xy}$ at different $T$ and 3.95~GPa; minor deviations from a linear behavior appear only at high fields. b) Almost linear $H$-dependence of the \textsc{Nernst} signal $N$ at different $T$ and 3.95~GPa. c) Contrasting field dependence $N(H)$ in NCCO (from ref.~\onlinecite{Wang_2003}) and CeCu$_2$Si$_2$ (3.95~GPa) at comparable $T/T_{c0}$ and $H_{c2}$. While a strong contribution from moving vortices leads to an increase of $N_{\text{NCCO}}$ below $H_{c2}$, $N_{\textnormal{CeCu$_2$Si$_2$}}(H)$ simply follows $\rho(H)$ and vanishes. d) The typical transition width of $N(H)$ at $H_{c2}$ roughly corresponds to the applied thermal gradient of $5-10\%$, as shown at $\sim 2$~K and 4.3~GPa. (lines are guides to the eye only)
\label{4timesB}}
\end{figure}

Last but not least let us comment on the transverse thermoelectric voltage in the presence of a perpendicular magnetic field, the \textsc{Nernst} effect (figs.~\ref{multi}, \ref{N_T} and \ref{4timesB}b). In a semi-classical, single-band \textsc{Boltzmann} picture, the quasiparticle contribution to the \textsc{Nernst} signal relies on the energy dependence of the relaxation time \cite{Oganesyan_2004,Kamran_2009}. Indeed, the appearance of a transverse electric field has its origin in the different relaxation times of the charge carriers involved in the heat current (induced by the applied $T$ gradient) and the charge carriers responsible of the compensating counter-flow. In ordinary metals, the relaxation time depends only weakly on energy, so that the \textsc{Nernst} signal is usually quite small (so-called \textsc{Sondheimer} cancelation \cite{Sondheimer_1948}). Assuming a linear energy dependence around the \textsc{Fermi} level, a simplified expression for the \textsc{Nernst} coefficient $\nu=N/B$ can be found \cite{Kamran_2009}: $\nu/T=\tfrac {\pi^2k_B}{3e}(=283[\mu\text{V}/\text{K}])\cdot\tfrac {\mu}{E_F/k_B}$ ($\mu$: carrier mobility). Hence a high mobility or a low \textsc{Fermi} energy can naturally lead to an enhanced \textsc{Nernst} signal, the latter case being often realized in HF compounds \cite{Kamran_2009}, for example in CeCoIn$_5$ (refs.~\onlinecite{Bel_2004,Onose_2007,Koichi_2007}). However, in CeCu$_2$Si$_2$ at ambient $p$, we only observe a tiny $N<0.2\mu$V/K at low $T$ and under 8~T (fig.~\ref{N_T} and ref.~\onlinecite{Rueetschi_2011}). A sign-change is observed below 1~K, likely related to the $A$-phase, and around 4~K, like in magnetoresistance. Here we are particularly interested in a possible signature of the $4f$ electron delocalization around $p_V$. In this respect, as a function of $p$, the most intriguing feature is a strong enhancement towards $p_V$, leading to a pronounced peak that coincides with the maximum in $T_c(p)$ (figs.~\ref{multi} and \ref{N_T}). Its value is more than one order of magnitude higher than at $p=0$ and roughly corresponds to half of the giant $N$ measured in CeCoIn$_5$ \cite{Bel_2004}. In addition, $N(T)$ exhibits a maximum around 2~K for $p<5$~GPa and independent of $H$, most likely revealing a characteristic energy scale (fig.~\ref{N_T}). For $p\sim p_V$, this assumption is reinforced by a negative peak also around 2~K in $S(T,H\neq 0)$ (fig.~\ref{TEP_T}). For $p>p_V$, $N$ behaves in a more ``conventional'' manner (smaller values, roughly $T$-linear), and it changes sign to negative. Let us add that generally a $H$-linear \textsc{Nernst} voltage is observed, as exemplified at 4~GPa in fig.~\ref{4timesB}b. We have also investigated the above relationship between $\nu/T$ and $\tfrac {\mu}{E_F/k_B}$ in the low $T$ limit \cite{Kamran_2009} (note that solely rough estimations for $E_F$ are available): CeCu$_2$Si$_2$ approaches the universal curve only at $p_V$, next to CeRu$_2$Si$_2$ (which may not be surprising since transport data show similarities \cite{Amato_1989}). However, away from $p_V$ (especially for $p\rightarrow 0$), the tiny $N$ inevitably drives the corresponding data quite off the universal line, yielding a slope different from 283[$\mu$V/K].

Apart from the quasi-particles, moving vortices in the mixed phase of a superconductor can also give rise to a transverse voltage \cite{Solomon_1967,Huebener_1969}, which is directly related to their velocity. This phenomenon has been extensively studied in the high $T_c$ superconductors, known to exhibit a strong enhancement of $N$ in the vortex-liquid state \cite{Ri_1994,Ong_2000,Wang_2003}. In contrast, no sizable vortex contribution to $N$ has yet been detected in HF systems (see for example ref.~\onlinecite{Bel_2004}), i.e. the vortices are assumed to be immobile. CeCu$_2$Si$_2$ does not deviate from this trend. In fig.~\ref{4timesB}c we compare the field dependence of the \textsc{Nernst} signal in NCCO (from ref.~\onlinecite{Wang_2003}) and CeCu$_2$Si$_2$ at a similar ratio of $T/T_c^{H=0}\sim 0.8$ (the $T\rightarrow 0$ limit of $H_{c2}$ is of the same order of magnitude in both compounds). While in CeCu$_2$Si$_2$ $N(H)$ simply follows the vanishing $\rho(H)$ (taking into account the applied thermal gradient of $5-10\%$ responsible for the transition width, fig.~\ref{4timesB}d), the \textsc{Nernst} signal of NCCO increases within the mixed state, due to a significant vortex contribution.

\begin{figure}[h]
\includegraphics[width=\linewidth]{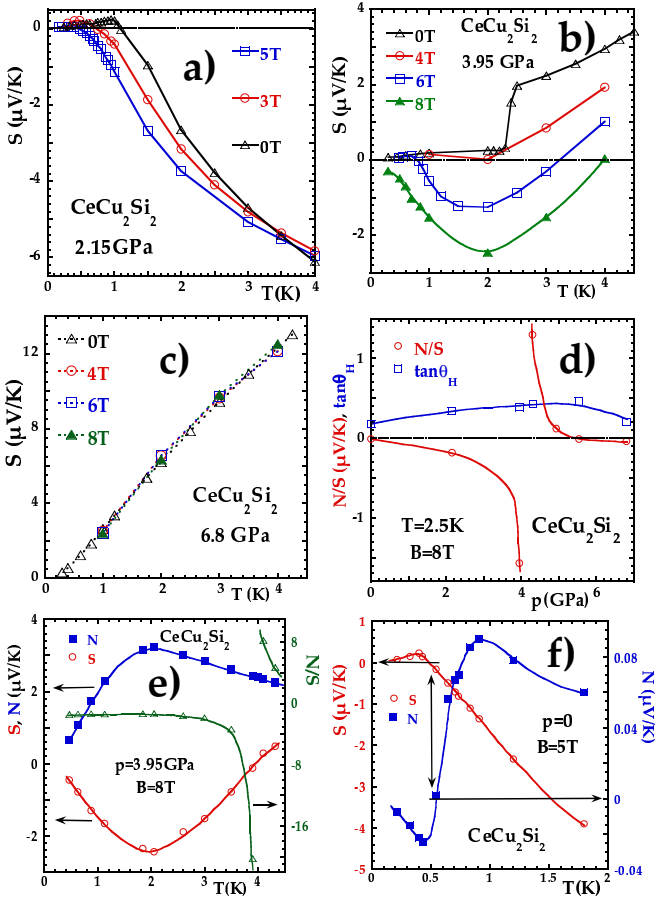}
\caption{a) to c) $p$-dependence of the sensitivity on magnetic field of the thermopower signal in CeCu$_2$Si$_2$, varying from weak at low $p$ over strong around $p_V$ to absent at high $p$. d) While the thermoelectric ratio $N/S=\tan\theta_{\text{TE}}$ exhibits an anomaly around 4~GPa, the electric analogue $\tan\theta_{\text{H}}$ displays a smooth $p$-dependence. e) This ``divergence'' at 4~GPa, a consequence of the signchange of thermopower (open circles), can also be observed in the $T$-dependence $N/S(T)$ (open triangles, right-hand scale). f) In other cases, like at $p=0$ and crossing the $A$ phase boundary, $S$ and $N$ change sign simultaneously, preventing any ``divergence''. (lines are guides to the eye only)
\label{6times}}
\end{figure}

Another interesting feature appears when comparing the magneto-thermal electric fields $E_x$ (longitudinal, probed by $S$) and $E_y$ (transverse, probed by $N$), see fig.~\ref{fig2}a. For some specific points in the $T-p-H$-space, particularly in vicinity to $p_V$ and under high magnetic field, $E_x$ vanishes (sign-change as already discussed above), whereas the transverse field $E_y$ stays finite. Such a situation is exemplified at 2.5~K and 8~T in fig.~\ref{6times}d, where the ratio $N/S(=\tan\theta_{\text{TE}})$ ``diverges'' as a function of $p$ around 4~GPa, and it is also seen in the $T$-dependence of $N/S$, shown at 4~GPa in fig.~\ref{6times}e. In other words, the electric field produced by a longitudinal heat current becomes exclusively transverse. Whether this happens accidentally and is simply related to the fact that $S$ and $N$ probe quite different phenomena in a complex multiband system evolving under $p$, or whether the orthogonal electric field occurs for more profound reasons, possibly linked to the $4f$ electron delocalization at $p_V$, is left for further studies. At this stage, we can only make the following comments: a similar trend has been reported in CeCoIn$_5$ in the low $H$ and $T$ limit \cite{Bel_2004} (but it is partly masked by SC). There a strongly energy-dependent elastic scattering rate had been invoked to explain the low \textsc{Seebeck} and large \textsc{Nernst} coefficients. Returning back to CeCu$_2$Si$_2$, in contrast to the ``divergence'' in $N/S$, no anomaly is detected in the $p$-dependence of the purely electric analogue $\rho_{xy}/\rho(=\tan\theta_{\text{H}})$, as also shown in fig.~\ref{6times}d and which may be trivial since $\rho$ can not change sign. Simultaneous measurements of $S$ and $N$ are not yet common in Ce compounds, and in the rare other reported cases (see for example refs.~\onlinecite{Pourret_2006} and \onlinecite{Johannsen_2008}), or like in CeCu$_2$Si$_2$ at $p=0$, $B=5~$T and $550~$mK (conditions corresponding to the transition into the $A$ phase), the signchange of $S$ coincides with a vanishing $N$ (illustrated in fig.~\ref{6times}f), so that no ``divergence'' occurs.

\section{Conclusion\label{}}

In conclusion, we have performed for the first time a multiprobe study under high pressure on a CeCu$_2$Si$_2$ single crystal with a unprecedented high superconducting $T_c$ and low $\rho_0$ in order to deepen the understanding of the expected Ce $4f$ electron delocalization at 4~GPa. A careful resistivity analysis unveils the proximity of CeCu$_2$Si$_2$ to the critical end point of the valence transition line, located at $p\sim4.5$~GPa and a slightly negative temperature. This means that CeCu$_2$Si$_2$ lies entirely in the valence cross-over regime and does not exhibit a first-order valence transition down to lowest temperatures. However, around $p_V$ and below about 30~K the complete $\rho(p)$ data set is governed by this adjacent critical end point, leading to a noteworthy data collapse. In addition to the critical end point, the experimentally determined valence cross-over line has been added as an important feature to the $p-T$ phase diagram of CeCu$_2$Si$_2$ (fig.~\ref{fig1}).

As an extraordinary tool, the new multiprobe setup (fig.~\ref{fig2}a) gives access to a huge parameter space, in which diverse physical quantities like $\rho$, $C^{ac}$, $\rho_{xy}$, $S$ and $N$ can be simultaneously compared on a single sample and under extreme conditions (low temperatures, high pressures and magnetic field). It should stimulate the detailed exploration of a variety of systems in strongly correlated matter. In CeCu$_2$Si$_2$, apart from resistivity, $\rho_{xy}$, $S$ and $N$ may also be considerably affected by the underlying critical end point since they reveal pronounced anomalies close to $p_V$ (fig.~\ref{multi}), highlighting an important change in the ground state properties of the system. It still needs to be elucidated whether these features are solely linked to the Ce $4f$ electron delocalization, or rather specific to the case where valence fluctuation-mediated superconductivity is present, like in CeCu$_2$Si$_2$. An extended theoretical framework and more experimental data under pressure from different probes will help to better identify these complex signatures and their relationship to the electron delocalization.

Overall, this study brings the scenario of superconductivity mediated by valence fluctuations to the front in heavy fermion physics, and further studies may reveal its wide-spread relevance, \emph{even in the more common case where $p_c$ and $p_V$ are not well separated}.


%



\begin{acknowledgments}
We acknowledge fruitful discussions with K. Behnia and J. Flouquet, technical assistance from M. Lopes and financial support from the Swiss National Science Foundation.
\end{acknowledgments}


\begin{thebibliography}{70}%
\makeatletter
\providecommand \@ifxundefined [1]{%
 \@ifx{#1\undefined}
}%
\providecommand \@ifnum [1]{%
 \ifnum #1\expandafter \@firstoftwo
 \else \expandafter \@secondoftwo
 \fi
}%
\providecommand \@ifx [1]{%
 \ifx #1\expandafter \@firstoftwo
 \else \expandafter \@secondoftwo
 \fi
}%
\providecommand \natexlab [1]{#1}%
\providecommand \enquote  [1]{``#1''}%
\providecommand \bibnamefont  [1]{#1}%
\providecommand \bibfnamefont [1]{#1}%
\providecommand \citenamefont [1]{#1}%
\providecommand \href@noop [0]{\@secondoftwo}%
\providecommand \href [0]{\begingroup \@sanitize@url \@href}%
\providecommand \@href[1]{\@@startlink{#1}\@@href}%
\providecommand \@@href[1]{\endgroup#1\@@endlink}%
\providecommand \@sanitize@url [0]{\catcode `\\12\catcode `\$12\catcode
  `\&12\catcode `\#12\catcode `\^12\catcode `\_12\catcode `\%12\relax}%
\providecommand \@@startlink[1]{}%
\providecommand \@@endlink[0]{}%
\providecommand \url  [0]{\begingroup\@sanitize@url \@url }%
\providecommand \@url [1]{\endgroup\@href {#1}{\urlprefix }}%
\providecommand \urlprefix  [0]{URL }%
\providecommand \Eprint [0]{\href }%
\providecommand \doibase [0]{http://dx.doi.org/}%
\providecommand \selectlanguage [0]{\@gobble}%
\providecommand \bibinfo  [0]{\@secondoftwo}%
\providecommand \bibfield  [0]{\@secondoftwo}%
\providecommand \translation [1]{[#1]}%
\providecommand \BibitemOpen [0]{}%
\providecommand \bibitemStop [0]{}%
\providecommand \bibitemNoStop [0]{.\EOS\space}%
\providecommand \EOS [0]{\spacefactor3000\relax}%
\providecommand \BibitemShut  [1]{\csname bibitem#1\endcsname}%
\let\auto@bib@innerbib\@empty
\bibitem [{\citenamefont {Steglich}\ \emph {et~al.}(1979)\citenamefont
  {Steglich}, \citenamefont {Aarts}, \citenamefont {Bredl}, \citenamefont
  {Lieke}, \citenamefont {Meschede}, \citenamefont {Franz},\ and\ \citenamefont
  {Sch\"afer}}]{Steglich_1979}%
  \BibitemOpen
  \bibfield  {author} {\bibinfo {author} {\bibfnamefont {F.}~\bibnamefont
  {Steglich}}, \bibinfo {author} {\bibfnamefont {J.}~\bibnamefont {Aarts}},
  \bibinfo {author} {\bibfnamefont {C.~D.}\ \bibnamefont {Bredl}}, \bibinfo
  {author} {\bibfnamefont {W.}~\bibnamefont {Lieke}}, \bibinfo {author}
  {\bibfnamefont {D.}~\bibnamefont {Meschede}}, \bibinfo {author}
  {\bibfnamefont {W.}~\bibnamefont {Franz}}, \ and\ \bibinfo {author}
  {\bibfnamefont {H.}~\bibnamefont {Sch\"afer}},\ }\href {\doibase
  10.1103/PhysRevLett.43.1892} {\bibfield  {journal} {\bibinfo  {journal}
  {Phys. Rev. Lett.}\ }\textbf {\bibinfo {volume} {43}},\ \bibinfo {pages}
  {1892} (\bibinfo {year} {1979})}\BibitemShut {NoStop}%
\bibitem [{\citenamefont {Jaccard}\ \emph {et~al.}(1985)\citenamefont
  {Jaccard}, \citenamefont {Mignot}, \citenamefont {Bellarbi}, \citenamefont
  {Benoit}, \citenamefont {Braun},\ and\ \citenamefont {Sierro}}]{Didier_1985}%
  \BibitemOpen
  \bibfield  {author} {\bibinfo {author} {\bibfnamefont {D.}~\bibnamefont
  {Jaccard}}, \bibinfo {author} {\bibfnamefont {J.}~\bibnamefont {Mignot}},
  \bibinfo {author} {\bibfnamefont {B.}~\bibnamefont {Bellarbi}}, \bibinfo
  {author} {\bibfnamefont {A.}~\bibnamefont {Benoit}}, \bibinfo {author}
  {\bibfnamefont {H.}~\bibnamefont {Braun}}, \ and\ \bibinfo {author}
  {\bibfnamefont {J.}~\bibnamefont {Sierro}},\ }\href {\doibase
  10.1016/0304-8853(85)90348-8} {\bibfield  {journal} {\bibinfo  {journal} {J.
  Magn. Magn. Mat.}\ }\textbf {\bibinfo {volume} {47\&48}},\ \bibinfo {pages}
  {23 } (\bibinfo {year} {1985})}\BibitemShut {NoStop}%
\bibitem [{\citenamefont {Jaccard}\ \emph {et~al.}(1999)\citenamefont
  {Jaccard}, \citenamefont {Wilhelm}, \citenamefont {Alami-Yadri},\ and\
  \citenamefont {Vargoz}}]{Didier_1999}%
  \BibitemOpen
  \bibfield  {author} {\bibinfo {author} {\bibfnamefont {D.}~\bibnamefont
  {Jaccard}}, \bibinfo {author} {\bibfnamefont {H.}~\bibnamefont {Wilhelm}},
  \bibinfo {author} {\bibfnamefont {K.}~\bibnamefont {Alami-Yadri}}, \ and\
  \bibinfo {author} {\bibfnamefont {E.}~\bibnamefont {Vargoz}},\ }\href
  {\doibase 10.1016/S0921-4526(98)01069-2} {\bibfield  {journal} {\bibinfo
  {journal} {Physica B}\ }\textbf {\bibinfo {volume} {259-261}},\ \bibinfo
  {pages} {1 } (\bibinfo {year} {1999})}\BibitemShut {NoStop}%
\bibitem [{\citenamefont {Holmes}\ \emph {et~al.}(2004)\citenamefont {Holmes},
  \citenamefont {Jaccard},\ and\ \citenamefont {Miyake}}]{Alex_PRB_2004}%
  \BibitemOpen
  \bibfield  {author} {\bibinfo {author} {\bibfnamefont {A.~T.}\ \bibnamefont
  {Holmes}}, \bibinfo {author} {\bibfnamefont {D.}~\bibnamefont {Jaccard}}, \
  and\ \bibinfo {author} {\bibfnamefont {K.}~\bibnamefont {Miyake}},\ }\href
  {\doibase 10.1103/PhysRevB.69.024508} {\bibfield  {journal} {\bibinfo
  {journal} {Phys. Rev. B}\ }\textbf {\bibinfo {volume} {69}},\ \bibinfo
  {pages} {024508} (\bibinfo {year} {2004})}\BibitemShut {NoStop}%
\bibitem [{\citenamefont {Holmes}\ \emph {et~al.}(2007)\citenamefont {Holmes},
  \citenamefont {Jaccard},\ and\ \citenamefont {Miyake}}]{Alex_2007}%
  \BibitemOpen
  \bibfield  {author} {\bibinfo {author} {\bibfnamefont {A.~T.}\ \bibnamefont
  {Holmes}}, \bibinfo {author} {\bibfnamefont {D.}~\bibnamefont {Jaccard}}, \
  and\ \bibinfo {author} {\bibfnamefont {K.}~\bibnamefont {Miyake}},\ }\href
  {\doibase 10.1143/JPSJ.76.051002} {\bibfield  {journal} {\bibinfo  {journal}
  {J. Phys. Soc. Jpn.}\ }\textbf {\bibinfo {volume} {76}},\ \bibinfo {pages}
  {051002} (\bibinfo {year} {2007})}\BibitemShut {NoStop}%
\bibitem [{\citenamefont {Miyake}(2007)}]{Miyake_2007}%
  \BibitemOpen
  \bibfield  {author} {\bibinfo {author} {\bibfnamefont {K.}~\bibnamefont
  {Miyake}},\ }\href {http://stacks.iop.org/0953-8984/19/i=12/a=125201}
  {\bibfield  {journal} {\bibinfo  {journal} {J. Phys.: Condens. Matter}\
  }\textbf {\bibinfo {volume} {19}},\ \bibinfo {pages} {125201} (\bibinfo
  {year} {2007})}\BibitemShut {NoStop}%
\bibitem [{\citenamefont {Watanabe}\ and\ \citenamefont
  {Miyake}(2011)}]{Wata_2011}%
  \BibitemOpen
  \bibfield  {author} {\bibinfo {author} {\bibfnamefont {S.}~\bibnamefont
  {Watanabe}}\ and\ \bibinfo {author} {\bibfnamefont {K.}~\bibnamefont
  {Miyake}},\ }\href {http://stacks.iop.org/0953-8984/23/i=9/a=094217}
  {\bibfield  {journal} {\bibinfo  {journal} {J. Phys.: Condens. Matter}\
  }\textbf {\bibinfo {volume} {23}},\ \bibinfo {pages} {094217} (\bibinfo
  {year} {2011})}\BibitemShut {NoStop}%
\bibitem [{\citenamefont {Jayaraman}(1965)}]{Jayaraman_1965}%
  \BibitemOpen
  \bibfield  {author} {\bibinfo {author} {\bibfnamefont {A.}~\bibnamefont
  {Jayaraman}},\ }\href {\doibase 10.1103/PhysRev.137.A179} {\bibfield
  {journal} {\bibinfo  {journal} {Phys. Rev.}\ }\textbf {\bibinfo {volume}
  {137}},\ \bibinfo {pages} {A179} (\bibinfo {year} {1965})}\BibitemShut
  {NoStop}%
\bibitem [{\citenamefont {Lipp}\ \emph {et~al.}(2008)\citenamefont {Lipp},
  \citenamefont {Jackson}, \citenamefont {Cynn}, \citenamefont {Aracne},
  \citenamefont {Evans},\ and\ \citenamefont {McMahan}}]{Lipp_2008}%
  \BibitemOpen
  \bibfield  {author} {\bibinfo {author} {\bibfnamefont {M.~J.}\ \bibnamefont
  {Lipp}}, \bibinfo {author} {\bibfnamefont {D.}~\bibnamefont {Jackson}},
  \bibinfo {author} {\bibfnamefont {H.}~\bibnamefont {Cynn}}, \bibinfo {author}
  {\bibfnamefont {C.}~\bibnamefont {Aracne}}, \bibinfo {author} {\bibfnamefont
  {W.~J.}\ \bibnamefont {Evans}}, \ and\ \bibinfo {author} {\bibfnamefont
  {A.~K.}\ \bibnamefont {McMahan}},\ }\href {\doibase
  10.1103/PhysRevLett.101.165703} {\bibfield  {journal} {\bibinfo  {journal}
  {Phys. Rev. Lett.}\ }\textbf {\bibinfo {volume} {101}},\ \bibinfo {pages}
  {165703} (\bibinfo {year} {2008})}\BibitemShut {NoStop}%
\bibitem [{\citenamefont {Vargoz}\ \emph {et~al.}(1998)\citenamefont {Vargoz},
  \citenamefont {Jaccard}, \citenamefont {Genoud}, \citenamefont {Brison},\
  and\ \citenamefont {Flouquet}}]{Vargoz_1998}%
  \BibitemOpen
  \bibfield  {author} {\bibinfo {author} {\bibfnamefont {E.}~\bibnamefont
  {Vargoz}}, \bibinfo {author} {\bibfnamefont {D.}~\bibnamefont {Jaccard}},
  \bibinfo {author} {\bibfnamefont {J.}~\bibnamefont {Genoud}}, \bibinfo
  {author} {\bibfnamefont {J.-P.}\ \bibnamefont {Brison}}, \ and\ \bibinfo
  {author} {\bibfnamefont {J.}~\bibnamefont {Flouquet}},\ }\href {\doibase
  10.1016/S0038-1098(98)00086-6} {\bibfield  {journal} {\bibinfo  {journal}
  {Solid State Commun.}\ }\textbf {\bibinfo {volume} {106}},\ \bibinfo {pages}
  {631 } (\bibinfo {year} {1998})}\BibitemShut {NoStop}%
\bibitem [{\citenamefont {Holmes}(2004)}]{Alex_thesis}%
  \BibitemOpen
  \bibfield  {author} {\bibinfo {author} {\bibfnamefont {A.~T.}\ \bibnamefont
  {Holmes}},\ }\emph {\bibinfo {title} {Exotic Superconducting Mechanisms in Fe
  and CeCu$_2$Si$_2$ under Pressure}},\ \href@noop {} {Ph.D. thesis},\ \bibinfo
   {school} {Universit\'e de Gen\`eve} (\bibinfo {year} {2004}),\ \Eprint
  {http://arxiv.org/abs/http://archive-ouverte.unige.ch/unige:284}
  {http://archive-ouverte.unige.ch/unige:284} \BibitemShut {NoStop}%
\bibitem [{\citenamefont {R\"uetschi}\ \emph {et~al.}(2011)\citenamefont
  {R\"uetschi}, \citenamefont {Sengupta}, \citenamefont {Seyfarth},\ and\
  \citenamefont {Jaccard}}]{Rueetschi_2011}%
  \BibitemOpen
  \bibfield  {author} {\bibinfo {author} {\bibfnamefont {A.-S.}\ \bibnamefont
  {R\"uetschi}}, \bibinfo {author} {\bibfnamefont {K.}~\bibnamefont
  {Sengupta}}, \bibinfo {author} {\bibfnamefont {G.}~\bibnamefont {Seyfarth}},
  \ and\ \bibinfo {author} {\bibfnamefont {D.}~\bibnamefont {Jaccard}},\ }\href
  {http://stacks.iop.org/1742-6596/273/i=1/a=012052} {\bibfield  {journal}
  {\bibinfo  {journal} {J. Phys.: Conf. Ser.}\ }\textbf {\bibinfo {volume}
  {273}},\ \bibinfo {pages} {012052} (\bibinfo {year} {2011})}\BibitemShut
  {NoStop}%
\bibitem [{\citenamefont {R\"uetschi}\ and\ \citenamefont
  {Jaccard}(2007)}]{Rueetschi_2007}%
  \BibitemOpen
  \bibfield  {author} {\bibinfo {author} {\bibfnamefont {A.-S.}\ \bibnamefont
  {R\"uetschi}}\ and\ \bibinfo {author} {\bibfnamefont {D.}~\bibnamefont
  {Jaccard}},\ }\href {\doibase 10.1063/1.2818788} {\bibfield  {journal}
  {\bibinfo  {journal} {Rev. Sci. Instrum.}\ }\textbf {\bibinfo {volume}
  {78}},\ \bibinfo {eid} {123901} (\bibinfo {year} {2007})}\BibitemShut
  {NoStop}%
\bibitem [{\citenamefont {Wang}\ \emph {et~al.}(2003)\citenamefont {Wang},
  \citenamefont {Ono}, \citenamefont {Onose}, \citenamefont {Gu}, \citenamefont
  {Ando}, \citenamefont {Tokura}, \citenamefont {Uchida},\ and\ \citenamefont
  {Ong}}]{Wang_2003}%
  \BibitemOpen
  \bibfield  {author} {\bibinfo {author} {\bibfnamefont {Y.}~\bibnamefont
  {Wang}}, \bibinfo {author} {\bibfnamefont {S.}~\bibnamefont {Ono}}, \bibinfo
  {author} {\bibfnamefont {Y.}~\bibnamefont {Onose}}, \bibinfo {author}
  {\bibfnamefont {G.}~\bibnamefont {Gu}}, \bibinfo {author} {\bibfnamefont
  {Y.}~\bibnamefont {Ando}}, \bibinfo {author} {\bibfnamefont {Y.}~\bibnamefont
  {Tokura}}, \bibinfo {author} {\bibfnamefont {S.}~\bibnamefont {Uchida}}, \
  and\ \bibinfo {author} {\bibfnamefont {N.~P.}\ \bibnamefont {Ong}},\ }\href
  {\doibase 10.1126/science.1078422} {\bibfield  {journal} {\bibinfo  {journal}
  {Science}\ }\textbf {\bibinfo {volume} {299}},\ \bibinfo {pages} {86}
  (\bibinfo {year} {2003})}\BibitemShut {NoStop}%
\bibitem [{\citenamefont {Wang}\ \emph {et~al.}(2006)\citenamefont {Wang},
  \citenamefont {Li},\ and\ \citenamefont {Ong}}]{Ong_2006}%
  \BibitemOpen
  \bibfield  {author} {\bibinfo {author} {\bibfnamefont {Y.}~\bibnamefont
  {Wang}}, \bibinfo {author} {\bibfnamefont {L.}~\bibnamefont {Li}}, \ and\
  \bibinfo {author} {\bibfnamefont {N.~P.}\ \bibnamefont {Ong}},\ }\href
  {\doibase 10.1103/PhysRevB.73.024510} {\bibfield  {journal} {\bibinfo
  {journal} {Phys. Rev. B}\ }\textbf {\bibinfo {volume} {73}},\ \bibinfo
  {pages} {024510} (\bibinfo {year} {2006})}\BibitemShut {NoStop}%
\bibitem [{\citenamefont {Blatt}(1957)}]{Blatt_1957}%
  \BibitemOpen
  \bibfield  {author} {\bibinfo {author} {\bibfnamefont {F.~J.}\ \bibnamefont
  {Blatt}},\ }\href {\doibase 10.1016/S0081-1947(08)60155-1} {\ \bibinfo
  {series} {Solid State Physics},\ \textbf {\bibinfo {volume} {4}},\ \bibinfo
  {pages} {199 } (\bibinfo {year} {1957})}\BibitemShut {NoStop}%
\bibitem [{\citenamefont {Steglich}\ \emph {et~al.}(1996)\citenamefont
  {Steglich}, \citenamefont {Gegenwart}, \citenamefont {Geibel}, \citenamefont
  {Helfrich}, \citenamefont {Hellmann}, \citenamefont {Lang}, \citenamefont
  {Link}, \citenamefont {Modler}, \citenamefont {Sparn}, \citenamefont
  {B\"uttgen},\ and\ \citenamefont {Loidl}}]{Steglich_1996}%
  \BibitemOpen
  \bibfield  {author} {\bibinfo {author} {\bibfnamefont {F.}~\bibnamefont
  {Steglich}}, \bibinfo {author} {\bibfnamefont {P.}~\bibnamefont {Gegenwart}},
  \bibinfo {author} {\bibfnamefont {C.}~\bibnamefont {Geibel}}, \bibinfo
  {author} {\bibfnamefont {R.}~\bibnamefont {Helfrich}}, \bibinfo {author}
  {\bibfnamefont {P.}~\bibnamefont {Hellmann}}, \bibinfo {author}
  {\bibfnamefont {M.}~\bibnamefont {Lang}}, \bibinfo {author} {\bibfnamefont
  {A.}~\bibnamefont {Link}}, \bibinfo {author} {\bibfnamefont {R.}~\bibnamefont
  {Modler}}, \bibinfo {author} {\bibfnamefont {G.}~\bibnamefont {Sparn}},
  \bibinfo {author} {\bibfnamefont {N.}~\bibnamefont {B\"uttgen}}, \ and\
  \bibinfo {author} {\bibfnamefont {A.}~\bibnamefont {Loidl}},\ }\href
  {\doibase 10.1016/0921-4526(96)00026-9} {\bibfield  {journal} {\bibinfo
  {journal} {Physica B}\ }\textbf {\bibinfo {volume} {223\&224}},\ \bibinfo
  {pages} {1 } (\bibinfo {year} {1996})}\BibitemShut {NoStop}%
\bibitem [{\citenamefont {Lengyel}\ \emph {et~al.}(2009)\citenamefont
  {Lengyel}, \citenamefont {Nicklas}, \citenamefont {Jeevan}, \citenamefont
  {Sparn}, \citenamefont {Geibel}, \citenamefont {Steglich}, \citenamefont
  {Yoshioka},\ and\ \citenamefont {Miyake}}]{Lengyel_2009}%
  \BibitemOpen
  \bibfield  {author} {\bibinfo {author} {\bibfnamefont {E.}~\bibnamefont
  {Lengyel}}, \bibinfo {author} {\bibfnamefont {M.}~\bibnamefont {Nicklas}},
  \bibinfo {author} {\bibfnamefont {H.~S.}\ \bibnamefont {Jeevan}}, \bibinfo
  {author} {\bibfnamefont {G.}~\bibnamefont {Sparn}}, \bibinfo {author}
  {\bibfnamefont {C.}~\bibnamefont {Geibel}}, \bibinfo {author} {\bibfnamefont
  {F.}~\bibnamefont {Steglich}}, \bibinfo {author} {\bibfnamefont
  {Y.}~\bibnamefont {Yoshioka}}, \ and\ \bibinfo {author} {\bibfnamefont
  {K.}~\bibnamefont {Miyake}},\ }\href {\doibase 10.1103/PhysRevB.80.140513}
  {\bibfield  {journal} {\bibinfo  {journal} {Phys. Rev. B}\ }\textbf {\bibinfo
  {volume} {80}},\ \bibinfo {pages} {140513} (\bibinfo {year}
  {2009})}\BibitemShut {NoStop}%
\bibitem [{\citenamefont {Bruls}\ \emph {et~al.}(1994)\citenamefont {Bruls},
  \citenamefont {Wolf}, \citenamefont {Finsterbusch}, \citenamefont
  {Thalmeier}, \citenamefont {Kouroudis}, \citenamefont {Sun}, \citenamefont
  {Assmus}, \citenamefont {L\"uthi}, \citenamefont {Lang}, \citenamefont
  {Gloos}, \citenamefont {Steglich},\ and\ \citenamefont
  {Modler}}]{Bruls_1994}%
  \BibitemOpen
  \bibfield  {author} {\bibinfo {author} {\bibfnamefont {G.}~\bibnamefont
  {Bruls}}, \bibinfo {author} {\bibfnamefont {B.}~\bibnamefont {Wolf}},
  \bibinfo {author} {\bibfnamefont {D.}~\bibnamefont {Finsterbusch}}, \bibinfo
  {author} {\bibfnamefont {P.}~\bibnamefont {Thalmeier}}, \bibinfo {author}
  {\bibfnamefont {I.}~\bibnamefont {Kouroudis}}, \bibinfo {author}
  {\bibfnamefont {W.}~\bibnamefont {Sun}}, \bibinfo {author} {\bibfnamefont
  {W.}~\bibnamefont {Assmus}}, \bibinfo {author} {\bibfnamefont
  {B.}~\bibnamefont {L\"uthi}}, \bibinfo {author} {\bibfnamefont
  {M.}~\bibnamefont {Lang}}, \bibinfo {author} {\bibfnamefont {K.}~\bibnamefont
  {Gloos}}, \bibinfo {author} {\bibfnamefont {F.}~\bibnamefont {Steglich}}, \
  and\ \bibinfo {author} {\bibfnamefont {R.}~\bibnamefont {Modler}},\ }\href
  {\doibase 10.1103/PhysRevLett.72.1754} {\bibfield  {journal} {\bibinfo
  {journal} {Phys. Rev. Lett.}\ }\textbf {\bibinfo {volume} {72}},\ \bibinfo
  {pages} {1754} (\bibinfo {year} {1994})}\BibitemShut {NoStop}%
\bibitem [{\citenamefont {Monthoux}\ and\ \citenamefont
  {Lonzarich}(2004)}]{Monthoux_2004}%
  \BibitemOpen
  \bibfield  {author} {\bibinfo {author} {\bibfnamefont {P.}~\bibnamefont
  {Monthoux}}\ and\ \bibinfo {author} {\bibfnamefont {G.~G.}\ \bibnamefont
  {Lonzarich}},\ }\href {\doibase 10.1103/PhysRevB.69.064517} {\bibfield
  {journal} {\bibinfo  {journal} {Phys. Rev. B}\ }\textbf {\bibinfo {volume}
  {69}},\ \bibinfo {pages} {064517} (\bibinfo {year} {2004})}\BibitemShut
  {NoStop}%
\bibitem [{\citenamefont {Rueff}\ \emph {et~al.}(2011)\citenamefont {Rueff},
  \citenamefont {Raymond}, \citenamefont {Taguchi}, \citenamefont {Sikora},
  \citenamefont {Iti\'e}, \citenamefont {Baudelet}, \citenamefont
  {Braithwaite}, \citenamefont {Knebel},\ and\ \citenamefont
  {Jaccard}}]{Rueff_2011}%
  \BibitemOpen
  \bibfield  {author} {\bibinfo {author} {\bibfnamefont {J.-P.}\ \bibnamefont
  {Rueff}}, \bibinfo {author} {\bibfnamefont {S.}~\bibnamefont {Raymond}},
  \bibinfo {author} {\bibfnamefont {M.}~\bibnamefont {Taguchi}}, \bibinfo
  {author} {\bibfnamefont {M.}~\bibnamefont {Sikora}}, \bibinfo {author}
  {\bibfnamefont {J.-P.}\ \bibnamefont {Iti\'e}}, \bibinfo {author}
  {\bibfnamefont {F.}~\bibnamefont {Baudelet}}, \bibinfo {author}
  {\bibfnamefont {D.}~\bibnamefont {Braithwaite}}, \bibinfo {author}
  {\bibfnamefont {G.}~\bibnamefont {Knebel}}, \ and\ \bibinfo {author}
  {\bibfnamefont {D.}~\bibnamefont {Jaccard}},\ }\href {\doibase
  10.1103/PhysRevLett.106.186405} {\bibfield  {journal} {\bibinfo  {journal}
  {Phys. Rev. Lett.}\ }\textbf {\bibinfo {volume} {106}},\ \bibinfo {pages}
  {186405} (\bibinfo {year} {2011})}\BibitemShut {NoStop}%
\bibitem [{\citenamefont {Fujiwara}\ \emph {et~al.}(2008)\citenamefont
  {Fujiwara}, \citenamefont {Hata}, \citenamefont {Kobayashi}, \citenamefont
  {Miyoshi}, \citenamefont {Takeuchi}, \citenamefont {Shimaoka}, \citenamefont
  {Kotegawa}, \citenamefont {Kobayashi}, \citenamefont {Geibel},\ and\
  \citenamefont {Steglich}}]{NQR_2008}%
  \BibitemOpen
  \bibfield  {author} {\bibinfo {author} {\bibfnamefont {K.}~\bibnamefont
  {Fujiwara}}, \bibinfo {author} {\bibfnamefont {Y.}~\bibnamefont {Hata}},
  \bibinfo {author} {\bibfnamefont {K.}~\bibnamefont {Kobayashi}}, \bibinfo
  {author} {\bibfnamefont {K.}~\bibnamefont {Miyoshi}}, \bibinfo {author}
  {\bibfnamefont {J.}~\bibnamefont {Takeuchi}}, \bibinfo {author}
  {\bibfnamefont {Y.}~\bibnamefont {Shimaoka}}, \bibinfo {author}
  {\bibfnamefont {H.}~\bibnamefont {Kotegawa}}, \bibinfo {author}
  {\bibfnamefont {T.~C.}\ \bibnamefont {Kobayashi}}, \bibinfo {author}
  {\bibfnamefont {C.}~\bibnamefont {Geibel}}, \ and\ \bibinfo {author}
  {\bibfnamefont {F.}~\bibnamefont {Steglich}},\ }\href {\doibase
  10.1143/JPSJ.77.123711} {\bibfield  {journal} {\bibinfo  {journal} {J. Phys.
  Soc. Jpn.}\ }\textbf {\bibinfo {volume} {77}},\ \bibinfo {pages} {123711}
  (\bibinfo {year} {2008})}\BibitemShut {NoStop}%
\bibitem [{\citenamefont {Yuan}\ \emph {et~al.}(2006)\citenamefont {Yuan},
  \citenamefont {Grosche}, \citenamefont {Deppe}, \citenamefont {Sparn},
  \citenamefont {Geibel},\ and\ \citenamefont {Steglich}}]{Yuan_2006}%
  \BibitemOpen
  \bibfield  {author} {\bibinfo {author} {\bibfnamefont {H.~Q.}\ \bibnamefont
  {Yuan}}, \bibinfo {author} {\bibfnamefont {F.~M.}\ \bibnamefont {Grosche}},
  \bibinfo {author} {\bibfnamefont {M.}~\bibnamefont {Deppe}}, \bibinfo
  {author} {\bibfnamefont {G.}~\bibnamefont {Sparn}}, \bibinfo {author}
  {\bibfnamefont {C.}~\bibnamefont {Geibel}}, \ and\ \bibinfo {author}
  {\bibfnamefont {F.}~\bibnamefont {Steglich}},\ }\href {\doibase
  10.1103/PhysRevLett.96.047008} {\bibfield  {journal} {\bibinfo  {journal}
  {Phys. Rev. Lett.}\ }\textbf {\bibinfo {volume} {96}},\ \bibinfo {pages}
  {047008} (\bibinfo {year} {2006})}\BibitemShut {NoStop}%
\bibitem [{\citenamefont {Yuan}\ \emph {et~al.}(2003)\citenamefont {Yuan},
  \citenamefont {Grosche}, \citenamefont {Deppe}, \citenamefont {Geibel},
  \citenamefont {Sparn},\ and\ \citenamefont {Steglich}}]{Yuan_2003}%
  \BibitemOpen
  \bibfield  {author} {\bibinfo {author} {\bibfnamefont {H.~Q.}\ \bibnamefont
  {Yuan}}, \bibinfo {author} {\bibfnamefont {F.~M.}\ \bibnamefont {Grosche}},
  \bibinfo {author} {\bibfnamefont {M.}~\bibnamefont {Deppe}}, \bibinfo
  {author} {\bibfnamefont {C.}~\bibnamefont {Geibel}}, \bibinfo {author}
  {\bibfnamefont {G.}~\bibnamefont {Sparn}}, \ and\ \bibinfo {author}
  {\bibfnamefont {F.}~\bibnamefont {Steglich}},\ }\href {\doibase
  10.1126/science.1091648} {\bibfield  {journal} {\bibinfo  {journal}
  {Science}\ }\textbf {\bibinfo {volume} {302}},\ \bibinfo {pages} {2104}
  (\bibinfo {year} {2003})}\BibitemShut {NoStop}%
\bibitem [{\citenamefont {Jaccard}\ and\ \citenamefont
  {Holmes}(2005)}]{Didier_2005}%
  \BibitemOpen
  \bibfield  {author} {\bibinfo {author} {\bibfnamefont {D.}~\bibnamefont
  {Jaccard}}\ and\ \bibinfo {author} {\bibfnamefont {A.~T.}\ \bibnamefont
  {Holmes}},\ }\href {\doibase 10.1016/j.physb.2005.01.056} {\bibfield
  {journal} {\bibinfo  {journal} {Physica B}\ }\textbf {\bibinfo {volume}
  {359-361}},\ \bibinfo {pages} {333 } (\bibinfo {year} {2005})}\BibitemShut
  {NoStop}%
\bibitem [{\citenamefont {Monthoux}\ \emph {et~al.}(2007)\citenamefont
  {Monthoux}, \citenamefont {Pines},\ and\ \citenamefont
  {Lonzarich}}]{Monthoux_2007}%
  \BibitemOpen
  \bibfield  {author} {\bibinfo {author} {\bibfnamefont {P.}~\bibnamefont
  {Monthoux}}, \bibinfo {author} {\bibfnamefont {D.}~\bibnamefont {Pines}}, \
  and\ \bibinfo {author} {\bibfnamefont {G.~G.}\ \bibnamefont {Lonzarich}},\
  }\href {\doibase 10.1038/nature06480} {\bibfield  {journal} {\bibinfo
  {journal} {Nature}\ }\textbf {\bibinfo {volume} {450}},\ \bibinfo {pages}
  {1177} (\bibinfo {year} {2007})}\BibitemShut {NoStop}%
\bibitem [{\citenamefont {Limelette}\ \emph {et~al.}(2003)\citenamefont
  {Limelette}, \citenamefont {Georges}, \citenamefont {J\'erome}, \citenamefont
  {Wzietek}, \citenamefont {Metcalf},\ and\ \citenamefont
  {Honig}}]{Limelette_2003}%
  \BibitemOpen
  \bibfield  {author} {\bibinfo {author} {\bibfnamefont {P.}~\bibnamefont
  {Limelette}}, \bibinfo {author} {\bibfnamefont {A.}~\bibnamefont {Georges}},
  \bibinfo {author} {\bibfnamefont {D.}~\bibnamefont {J\'erome}}, \bibinfo
  {author} {\bibfnamefont {P.}~\bibnamefont {Wzietek}}, \bibinfo {author}
  {\bibfnamefont {P.}~\bibnamefont {Metcalf}}, \ and\ \bibinfo {author}
  {\bibfnamefont {J.~M.}\ \bibnamefont {Honig}},\ }\href {\doibase
  10.1126/science.1088386} {\bibfield  {journal} {\bibinfo  {journal}
  {Science}\ }\textbf {\bibinfo {volume} {302}},\ \bibinfo {pages} {89}
  (\bibinfo {year} {2003})}\BibitemShut {NoStop}%
\bibitem [{\citenamefont {Lawrence}\ \emph {et~al.}(1975)\citenamefont
  {Lawrence}, \citenamefont {Croft},\ and\ \citenamefont
  {Parks}}]{Lawrence_1975}%
  \BibitemOpen
  \bibfield  {author} {\bibinfo {author} {\bibfnamefont {J.~M.}\ \bibnamefont
  {Lawrence}}, \bibinfo {author} {\bibfnamefont {M.~C.}\ \bibnamefont {Croft}},
  \ and\ \bibinfo {author} {\bibfnamefont {R.~D.}\ \bibnamefont {Parks}},\
  }\href {\doibase 10.1103/PhysRevLett.35.289} {\bibfield  {journal} {\bibinfo
  {journal} {Phys. Rev. Lett.}\ }\textbf {\bibinfo {volume} {35}},\ \bibinfo
  {pages} {289} (\bibinfo {year} {1975})}\BibitemShut {NoStop}%
\bibitem [{\citenamefont {Wilhelm}\ and\ \citenamefont
  {Jaccard}(2002)}]{Wilhelm_2002}%
  \BibitemOpen
  \bibfield  {author} {\bibinfo {author} {\bibfnamefont {H.}~\bibnamefont
  {Wilhelm}}\ and\ \bibinfo {author} {\bibfnamefont {D.}~\bibnamefont
  {Jaccard}},\ }\href {\doibase 10.1103/PhysRevB.66.064428} {\bibfield
  {journal} {\bibinfo  {journal} {Phys. Rev. B}\ }\textbf {\bibinfo {volume}
  {66}},\ \bibinfo {pages} {064428} (\bibinfo {year} {2002})}\BibitemShut
  {NoStop}%
\bibitem [{\citenamefont {Wilhelm}\ and\ \citenamefont
  {Jaccard}(2004)}]{Wilhelm_2004}%
  \BibitemOpen
  \bibfield  {author} {\bibinfo {author} {\bibfnamefont {H.}~\bibnamefont
  {Wilhelm}}\ and\ \bibinfo {author} {\bibfnamefont {D.}~\bibnamefont
  {Jaccard}},\ }\href {\doibase 10.1103/PhysRevB.69.214408} {\bibfield
  {journal} {\bibinfo  {journal} {Phys. Rev. B}\ }\textbf {\bibinfo {volume}
  {69}},\ \bibinfo {pages} {214408} (\bibinfo {year} {2004})}\BibitemShut
  {NoStop}%
\bibitem [{\citenamefont {{Wilhelm}}\ \emph {et~al.}()\citenamefont
  {{Wilhelm}}, \citenamefont {{Raymond}}, \citenamefont {{Jaccard}},
  \citenamefont {{Stockert}},\ and\ \citenamefont
  {{L\"ohneysen}}}]{Wilhelm_1999}%
  \BibitemOpen
  \bibfield  {author} {\bibinfo {author} {\bibfnamefont {H.}~\bibnamefont
  {{Wilhelm}}}, \bibinfo {author} {\bibfnamefont {S.}~\bibnamefont
  {{Raymond}}}, \bibinfo {author} {\bibfnamefont {D.}~\bibnamefont
  {{Jaccard}}}, \bibinfo {author} {\bibfnamefont {O.}~\bibnamefont
  {{Stockert}}}, \ and\ \bibinfo {author} {\bibfnamefont {H.~v.}\ \bibnamefont
  {{L\"ohneysen}}},\ }\href@noop {} {}\Eprint
  {http://arxiv.org/abs/cond-mat/9908442} {arXiv:cond-mat/9908442} \BibitemShut
  {NoStop}%
\bibitem [{\citenamefont {Muramatsu}\ \emph {et~al.}(2001)\citenamefont
  {Muramatsu}, \citenamefont {Tateiwa}, \citenamefont {Kobayashi},
  \citenamefont {Shimizu}, \citenamefont {Amaya}, \citenamefont {Aoki},
  \citenamefont {Shishido}, \citenamefont {Haga},\ and\ \citenamefont
  {\={O}nuki}}]{Muramatsu_2001}%
  \BibitemOpen
  \bibfield  {author} {\bibinfo {author} {\bibfnamefont {T.}~\bibnamefont
  {Muramatsu}}, \bibinfo {author} {\bibfnamefont {N.}~\bibnamefont {Tateiwa}},
  \bibinfo {author} {\bibfnamefont {T.~C.}\ \bibnamefont {Kobayashi}}, \bibinfo
  {author} {\bibfnamefont {K.}~\bibnamefont {Shimizu}}, \bibinfo {author}
  {\bibfnamefont {K.}~\bibnamefont {Amaya}}, \bibinfo {author} {\bibfnamefont
  {D.}~\bibnamefont {Aoki}}, \bibinfo {author} {\bibfnamefont {H.}~\bibnamefont
  {Shishido}}, \bibinfo {author} {\bibfnamefont {Y.}~\bibnamefont {Haga}}, \
  and\ \bibinfo {author} {\bibfnamefont {Y.}~\bibnamefont {\={O}nuki}},\ }\href
  {\doibase 10.1143/JPSJ.70.3362} {\bibfield  {journal} {\bibinfo  {journal}
  {J. Phys. Soc. Jpn.}\ }\textbf {\bibinfo {volume} {70}},\ \bibinfo {pages}
  {3362} (\bibinfo {year} {2001})}\BibitemShut {NoStop}%
\bibitem [{\citenamefont {Knebel}\ \emph {et~al.}(2008)\citenamefont {Knebel},
  \citenamefont {Aoki}, \citenamefont {Brison},\ and\ \citenamefont
  {Flouquet}}]{Knebel_2008}%
  \BibitemOpen
  \bibfield  {author} {\bibinfo {author} {\bibfnamefont {G.}~\bibnamefont
  {Knebel}}, \bibinfo {author} {\bibfnamefont {D.}~\bibnamefont {Aoki}},
  \bibinfo {author} {\bibfnamefont {J.-P.}\ \bibnamefont {Brison}}, \ and\
  \bibinfo {author} {\bibfnamefont {J.}~\bibnamefont {Flouquet}},\ }\href
  {\doibase 10.1143/JPSJ.77.114704} {\bibfield  {journal} {\bibinfo  {journal}
  {J. Phys. Soc. Jpn.}\ }\textbf {\bibinfo {volume} {77}},\ \bibinfo {pages}
  {114704} (\bibinfo {year} {2008})}\BibitemShut {NoStop}%
\bibitem [{\citenamefont {Park}\ \emph {et~al.}(2008)\citenamefont {Park},
  \citenamefont {Sidorov}, \citenamefont {Ronning}, \citenamefont {Zhu},
  \citenamefont {Tokiwa}, \citenamefont {Lee}, \citenamefont {Bauer},
  \citenamefont {Movshovich}, \citenamefont {Sarrao},\ and\ \citenamefont
  {Thompson}}]{Park_2008}%
  \BibitemOpen
  \bibfield  {author} {\bibinfo {author} {\bibfnamefont {T.}~\bibnamefont
  {Park}}, \bibinfo {author} {\bibfnamefont {V.~A.}\ \bibnamefont {Sidorov}},
  \bibinfo {author} {\bibfnamefont {F.}~\bibnamefont {Ronning}}, \bibinfo
  {author} {\bibfnamefont {J.-X.}\ \bibnamefont {Zhu}}, \bibinfo {author}
  {\bibfnamefont {Y.}~\bibnamefont {Tokiwa}}, \bibinfo {author} {\bibfnamefont
  {H.}~\bibnamefont {Lee}}, \bibinfo {author} {\bibfnamefont {E.~D.}\
  \bibnamefont {Bauer}}, \bibinfo {author} {\bibfnamefont {R.}~\bibnamefont
  {Movshovich}}, \bibinfo {author} {\bibfnamefont {J.~L.}\ \bibnamefont
  {Sarrao}}, \ and\ \bibinfo {author} {\bibfnamefont {J.~D.}\ \bibnamefont
  {Thompson}},\ }\href@noop {} {\bibfield  {journal} {\bibinfo  {journal}
  {Nature}\ }\textbf {\bibinfo {volume} {456}},\ \bibinfo {pages} {366}
  (\bibinfo {year} {2008})}\BibitemShut {NoStop}%
\bibitem [{\citenamefont {Vargoz}\ and\ \citenamefont
  {Jaccard}(1998)}]{Vargoz_1998_JMMM}%
  \BibitemOpen
  \bibfield  {author} {\bibinfo {author} {\bibfnamefont {E.}~\bibnamefont
  {Vargoz}}\ and\ \bibinfo {author} {\bibfnamefont {D.}~\bibnamefont
  {Jaccard}},\ }\href {\doibase 10.1016/S0304-8853(97)00688-4} {\bibfield
  {journal} {\bibinfo  {journal} {J. Magn. Magn. Mat.}\ }\textbf {\bibinfo
  {volume} {177-181}},\ \bibinfo {pages} {294 } (\bibinfo {year}
  {1998})}\BibitemShut {NoStop}%
\bibitem [{\citenamefont {Vargoz}(1998)}]{Vargoz_thesis}%
  \BibitemOpen
  \bibfield  {author} {\bibinfo {author} {\bibfnamefont {E.}~\bibnamefont
  {Vargoz}},\ }\emph {\bibinfo {title} {Propri\'et\'es de transport sous
  pression de compos\'es \`a fermions lourds supraconducteurs: CeCu$_2$,
  CeCu$_2$Ge$_2$ et CeCu$_2$Si$_2$.}},\ \href@noop {} {Ph.D. thesis},\ \bibinfo
   {school} {Universit\'e de Gen\`eve} (\bibinfo {year} {1998})\BibitemShut
  {NoStop}%
\bibitem [{\citenamefont {Capan}\ \emph {et~al.}(2004)\citenamefont {Capan},
  \citenamefont {Bianchi}, \citenamefont {Ronning}, \citenamefont {Lacerda},
  \citenamefont {Thompson}, \citenamefont {Hundley}, \citenamefont {Pagliuso},
  \citenamefont {Sarrao},\ and\ \citenamefont {Movshovich}}]{Capan_2004}%
  \BibitemOpen
  \bibfield  {author} {\bibinfo {author} {\bibfnamefont {C.}~\bibnamefont
  {Capan}}, \bibinfo {author} {\bibfnamefont {A.}~\bibnamefont {Bianchi}},
  \bibinfo {author} {\bibfnamefont {F.}~\bibnamefont {Ronning}}, \bibinfo
  {author} {\bibfnamefont {A.}~\bibnamefont {Lacerda}}, \bibinfo {author}
  {\bibfnamefont {J.~D.}\ \bibnamefont {Thompson}}, \bibinfo {author}
  {\bibfnamefont {M.~F.}\ \bibnamefont {Hundley}}, \bibinfo {author}
  {\bibfnamefont {P.~G.}\ \bibnamefont {Pagliuso}}, \bibinfo {author}
  {\bibfnamefont {J.~L.}\ \bibnamefont {Sarrao}}, \ and\ \bibinfo {author}
  {\bibfnamefont {R.}~\bibnamefont {Movshovich}},\ }\href {\doibase
  10.1103/PhysRevB.70.180502} {\bibfield  {journal} {\bibinfo  {journal} {Phys.
  Rev. B}\ }\textbf {\bibinfo {volume} {70}},\ \bibinfo {pages} {180502}
  (\bibinfo {year} {2004})}\BibitemShut {NoStop}%
\bibitem [{\citenamefont {Capan}\ \emph {et~al.}(2009)\citenamefont {Capan},
  \citenamefont {Balicas}, \citenamefont {Murphy}, \citenamefont {Palm},
  \citenamefont {Movshovich}, \citenamefont {Hall}, \citenamefont {Tozer},
  \citenamefont {Hundley}, \citenamefont {Bauer}, \citenamefont {Thompson},
  \citenamefont {Sarrao}, \citenamefont {DiTusa}, \citenamefont {Goodrich},\
  and\ \citenamefont {Fisk}}]{Capan_2009}%
  \BibitemOpen
  \bibfield  {author} {\bibinfo {author} {\bibfnamefont {C.}~\bibnamefont
  {Capan}}, \bibinfo {author} {\bibfnamefont {L.}~\bibnamefont {Balicas}},
  \bibinfo {author} {\bibfnamefont {T.~P.}\ \bibnamefont {Murphy}}, \bibinfo
  {author} {\bibfnamefont {E.~C.}\ \bibnamefont {Palm}}, \bibinfo {author}
  {\bibfnamefont {R.}~\bibnamefont {Movshovich}}, \bibinfo {author}
  {\bibfnamefont {D.}~\bibnamefont {Hall}}, \bibinfo {author} {\bibfnamefont
  {S.~W.}\ \bibnamefont {Tozer}}, \bibinfo {author} {\bibfnamefont {M.~F.}\
  \bibnamefont {Hundley}}, \bibinfo {author} {\bibfnamefont {E.~D.}\
  \bibnamefont {Bauer}}, \bibinfo {author} {\bibfnamefont {J.~D.}\ \bibnamefont
  {Thompson}}, \bibinfo {author} {\bibfnamefont {J.~L.}\ \bibnamefont
  {Sarrao}}, \bibinfo {author} {\bibfnamefont {J.~F.}\ \bibnamefont {DiTusa}},
  \bibinfo {author} {\bibfnamefont {R.~G.}\ \bibnamefont {Goodrich}}, \ and\
  \bibinfo {author} {\bibfnamefont {Z.}~\bibnamefont {Fisk}},\ }\href {\doibase
  10.1103/PhysRevB.80.094518} {\bibfield  {journal} {\bibinfo  {journal} {Phys.
  Rev. B}\ }\textbf {\bibinfo {volume} {80}},\ \bibinfo {pages} {094518}
  (\bibinfo {year} {2009})}\BibitemShut {NoStop}%
\bibitem [{\citenamefont {Sarrao}\ \emph {et~al.}(1999)\citenamefont {Sarrao},
  \citenamefont {Immer}, \citenamefont {Fisk}, \citenamefont {Booth},
  \citenamefont {Figueroa}, \citenamefont {Lawrence}, \citenamefont {Modler},
  \citenamefont {Cornelius}, \citenamefont {Hundley}, \citenamefont {Kwei},
  \citenamefont {Thompson},\ and\ \citenamefont {Bridges}}]{Sarrao_1999}%
  \BibitemOpen
  \bibfield  {author} {\bibinfo {author} {\bibfnamefont {J.~L.}\ \bibnamefont
  {Sarrao}}, \bibinfo {author} {\bibfnamefont {C.~D.}\ \bibnamefont {Immer}},
  \bibinfo {author} {\bibfnamefont {Z.}~\bibnamefont {Fisk}}, \bibinfo {author}
  {\bibfnamefont {C.~H.}\ \bibnamefont {Booth}}, \bibinfo {author}
  {\bibfnamefont {E.}~\bibnamefont {Figueroa}}, \bibinfo {author}
  {\bibfnamefont {J.~M.}\ \bibnamefont {Lawrence}}, \bibinfo {author}
  {\bibfnamefont {R.}~\bibnamefont {Modler}}, \bibinfo {author} {\bibfnamefont
  {A.~L.}\ \bibnamefont {Cornelius}}, \bibinfo {author} {\bibfnamefont {M.~F.}\
  \bibnamefont {Hundley}}, \bibinfo {author} {\bibfnamefont {G.~H.}\
  \bibnamefont {Kwei}}, \bibinfo {author} {\bibfnamefont {J.~D.}\ \bibnamefont
  {Thompson}}, \ and\ \bibinfo {author} {\bibfnamefont {F.}~\bibnamefont
  {Bridges}},\ }\href {\doibase 10.1103/PhysRevB.59.6855} {\bibfield  {journal}
  {\bibinfo  {journal} {Phys. Rev. B}\ }\textbf {\bibinfo {volume} {59}},\
  \bibinfo {pages} {6855} (\bibinfo {year} {1999})}\BibitemShut {NoStop}%
\bibitem [{\citenamefont {Wada}\ \emph {et~al.}(2008)\citenamefont {Wada},
  \citenamefont {Yamamoto}, \citenamefont {Ishida},\ and\ \citenamefont
  {Sarrao}}]{Wada_2008}%
  \BibitemOpen
  \bibfield  {author} {\bibinfo {author} {\bibfnamefont {S.}~\bibnamefont
  {Wada}}, \bibinfo {author} {\bibfnamefont {A.}~\bibnamefont {Yamamoto}},
  \bibinfo {author} {\bibfnamefont {K.}~\bibnamefont {Ishida}}, \ and\ \bibinfo
  {author} {\bibfnamefont {J.~L.}\ \bibnamefont {Sarrao}},\ }\href
  {http://stacks.iop.org/0953-8984/20/i=17/a=175201} {\bibfield  {journal}
  {\bibinfo  {journal} {J. Phys.: Condens. Matter}\ }\textbf {\bibinfo {volume}
  {20}},\ \bibinfo {pages} {175201} (\bibinfo {year} {2008})}\BibitemShut
  {NoStop}%
\bibitem [{\citenamefont {Thomas}\ \emph {et~al.}(1996)\citenamefont {Thomas},
  \citenamefont {Ayache}, \citenamefont {Fomine}, \citenamefont {Thomasson},\
  and\ \citenamefont {Geibel}}]{Thomas_1996}%
  \BibitemOpen
  \bibfield  {author} {\bibinfo {author} {\bibfnamefont {F.}~\bibnamefont
  {Thomas}}, \bibinfo {author} {\bibfnamefont {C.}~\bibnamefont {Ayache}},
  \bibinfo {author} {\bibfnamefont {I.~A.}\ \bibnamefont {Fomine}}, \bibinfo
  {author} {\bibfnamefont {J.}~\bibnamefont {Thomasson}}, \ and\ \bibinfo
  {author} {\bibfnamefont {C.}~\bibnamefont {Geibel}},\ }\href
  {http://stacks.iop.org/0953-8984/8/i=4/a=002} {\bibfield  {journal} {\bibinfo
   {journal} {J. Phys.: Condens. Matter}\ }\textbf {\bibinfo {volume} {8}},\
  \bibinfo {pages} {L51} (\bibinfo {year} {1996})}\BibitemShut {NoStop}%
\bibitem [{\citenamefont {Lengyel}\ \emph {et~al.}(2011)\citenamefont
  {Lengyel}, \citenamefont {Nicklas}, \citenamefont {Jeevan}, \citenamefont
  {Geibel},\ and\ \citenamefont {Steglich}}]{Lengyel_2011}%
  \BibitemOpen
  \bibfield  {author} {\bibinfo {author} {\bibfnamefont {E.}~\bibnamefont
  {Lengyel}}, \bibinfo {author} {\bibfnamefont {M.}~\bibnamefont {Nicklas}},
  \bibinfo {author} {\bibfnamefont {H.~S.}\ \bibnamefont {Jeevan}}, \bibinfo
  {author} {\bibfnamefont {C.}~\bibnamefont {Geibel}}, \ and\ \bibinfo {author}
  {\bibfnamefont {F.}~\bibnamefont {Steglich}},\ }\href {\doibase
  10.1103/PhysRevLett.107.057001} {\bibfield  {journal} {\bibinfo  {journal}
  {Phys. Rev. Lett.}\ }\textbf {\bibinfo {volume} {107}},\ \bibinfo {pages}
  {057001} (\bibinfo {year} {2011})}\BibitemShut {NoStop}%
\bibitem [{\citenamefont {Holmes}\ \emph {et~al.}(2005)\citenamefont {Holmes},
  \citenamefont {Jaccard}, \citenamefont {Jeevan}, \citenamefont {Geibel},\
  and\ \citenamefont {Ishikawa}}]{Alex_2005}%
  \BibitemOpen
  \bibfield  {author} {\bibinfo {author} {\bibfnamefont {A.~T.}\ \bibnamefont
  {Holmes}}, \bibinfo {author} {\bibfnamefont {D.}~\bibnamefont {Jaccard}},
  \bibinfo {author} {\bibfnamefont {H.~S.}\ \bibnamefont {Jeevan}}, \bibinfo
  {author} {\bibfnamefont {C.}~\bibnamefont {Geibel}}, \ and\ \bibinfo {author}
  {\bibfnamefont {M.}~\bibnamefont {Ishikawa}},\ }\href
  {http://stacks.iop.org/0953-8984/17/i=35/a=009} {\bibfield  {journal}
  {\bibinfo  {journal} {J. Phys.: Condens. Matter}\ }\textbf {\bibinfo {volume}
  {17}},\ \bibinfo {pages} {5423} (\bibinfo {year} {2005})}\BibitemShut
  {NoStop}%
\bibitem [{\citenamefont {Park}\ \emph {et~al.}(2012)\citenamefont {Park},
  \citenamefont {Lee}, \citenamefont {Martin}, \citenamefont {Lu},
  \citenamefont {Sidorov}, \citenamefont {Gofryk}, \citenamefont {Ronning},
  \citenamefont {Bauer},\ and\ \citenamefont {Thompson}}]{Park_2012}%
  \BibitemOpen
  \bibfield  {author} {\bibinfo {author} {\bibfnamefont {T.}~\bibnamefont
  {Park}}, \bibinfo {author} {\bibfnamefont {H.}~\bibnamefont {Lee}}, \bibinfo
  {author} {\bibfnamefont {I.}~\bibnamefont {Martin}}, \bibinfo {author}
  {\bibfnamefont {X.}~\bibnamefont {Lu}}, \bibinfo {author} {\bibfnamefont
  {V.~A.}\ \bibnamefont {Sidorov}}, \bibinfo {author} {\bibfnamefont
  {K.}~\bibnamefont {Gofryk}}, \bibinfo {author} {\bibfnamefont
  {F.}~\bibnamefont {Ronning}}, \bibinfo {author} {\bibfnamefont {E.~D.}\
  \bibnamefont {Bauer}}, \ and\ \bibinfo {author} {\bibfnamefont {J.~D.}\
  \bibnamefont {Thompson}},\ }\href {\doibase 10.1103/PhysRevLett.108.077003}
  {\bibfield  {journal} {\bibinfo  {journal} {Phys. Rev. Lett.}\ }\textbf
  {\bibinfo {volume} {108}},\ \bibinfo {pages} {077003} (\bibinfo {year}
  {2012})}\BibitemShut {NoStop}%
\bibitem [{\citenamefont {Rauchschwalbe}\ \emph {et~al.}(1987)\citenamefont
  {Rauchschwalbe}, \citenamefont {Steglich}, \citenamefont {de~Visser},\ and\
  \citenamefont {Franse}}]{Rauchschwalbe_1987}%
  \BibitemOpen
  \bibfield  {author} {\bibinfo {author} {\bibfnamefont {U.}~\bibnamefont
  {Rauchschwalbe}}, \bibinfo {author} {\bibfnamefont {F.}~\bibnamefont
  {Steglich}}, \bibinfo {author} {\bibfnamefont {A.}~\bibnamefont {de~Visser}},
  \ and\ \bibinfo {author} {\bibfnamefont {J.}~\bibnamefont {Franse}},\ }\href
  {\doibase 10.1016/0304-8853(87)90607-X} {\bibfield  {journal} {\bibinfo
  {journal} {J. Magn. Magn. Mat.}\ }\textbf {\bibinfo {volume} {63\&64}},\
  \bibinfo {pages} {347 } (\bibinfo {year} {1987})}\BibitemShut {NoStop}%
\bibitem [{\citenamefont {Aliev}\ \emph {et~al.}(1984)\citenamefont {Aliev},
  \citenamefont {Brandt}, \citenamefont {Moshchalkov},\ and\ \citenamefont
  {Chudinov}}]{Aliev_1984}%
  \BibitemOpen
  \bibfield  {author} {\bibinfo {author} {\bibfnamefont {F.~G.}\ \bibnamefont
  {Aliev}}, \bibinfo {author} {\bibfnamefont {N.~B.}\ \bibnamefont {Brandt}},
  \bibinfo {author} {\bibfnamefont {V.~V.}\ \bibnamefont {Moshchalkov}}, \ and\
  \bibinfo {author} {\bibfnamefont {S.~M.}\ \bibnamefont {Chudinov}},\ }\href
  {http://dx.doi.org/10.1007/BF00681517} {\bibfield  {journal} {\bibinfo
  {journal} {J. Low Temp. Phys.}\ }\textbf {\bibinfo {volume} {57}},\ \bibinfo
  {pages} {61} (\bibinfo {year} {1984})},\ \bibinfo {note}
  {10.1007/BF00681517}\BibitemShut {NoStop}%
\bibitem [{\citenamefont {Pippard}(1989)}]{Pippard_1989}%
  \BibitemOpen
  \bibfield  {author} {\bibinfo {author} {\bibfnamefont {A.}~\bibnamefont
  {Pippard}},\ }\href@noop {} {\emph {\bibinfo {title} {Magnetoresistance in
  Metals}}}\ (\bibinfo  {publisher} {Cambridge University Press},\ \bibinfo
  {address} {Cambridge},\ \bibinfo {year} {1989})\BibitemShut {NoStop}%
\bibitem [{\citenamefont {Nakajima}\ \emph {et~al.}(2007)\citenamefont
  {Nakajima}, \citenamefont {Shishido}, \citenamefont {Nakai}, \citenamefont
  {Shibauchi}, \citenamefont {Behnia}, \citenamefont {Izawa}, \citenamefont
  {Hedo}, \citenamefont {Uwatoko}, \citenamefont {Matsumoto}, \citenamefont
  {Settai}, \citenamefont {\={O}nuki}, \citenamefont {Kontani},\ and\
  \citenamefont {Matsuda}}]{Nakajima_2007}%
  \BibitemOpen
  \bibfield  {author} {\bibinfo {author} {\bibfnamefont {Y.}~\bibnamefont
  {Nakajima}}, \bibinfo {author} {\bibfnamefont {H.}~\bibnamefont {Shishido}},
  \bibinfo {author} {\bibfnamefont {H.}~\bibnamefont {Nakai}}, \bibinfo
  {author} {\bibfnamefont {T.}~\bibnamefont {Shibauchi}}, \bibinfo {author}
  {\bibfnamefont {K.}~\bibnamefont {Behnia}}, \bibinfo {author} {\bibfnamefont
  {K.}~\bibnamefont {Izawa}}, \bibinfo {author} {\bibfnamefont
  {M.}~\bibnamefont {Hedo}}, \bibinfo {author} {\bibfnamefont {Y.}~\bibnamefont
  {Uwatoko}}, \bibinfo {author} {\bibfnamefont {T.}~\bibnamefont {Matsumoto}},
  \bibinfo {author} {\bibfnamefont {R.}~\bibnamefont {Settai}}, \bibinfo
  {author} {\bibfnamefont {Y.}~\bibnamefont {\={O}nuki}}, \bibinfo {author}
  {\bibfnamefont {H.}~\bibnamefont {Kontani}}, \ and\ \bibinfo {author}
  {\bibfnamefont {Y.}~\bibnamefont {Matsuda}},\ }\href {\doibase
  10.1143/JPSJ.76.024703} {\bibfield  {journal} {\bibinfo  {journal} {J. Phys.
  Soc. Jpn.}\ }\textbf {\bibinfo {volume} {76}},\ \bibinfo {pages} {024703}
  (\bibinfo {year} {2007})}\BibitemShut {NoStop}%
\bibitem [{\citenamefont {\={O}nuki}\ \emph {et~al.}(1989)\citenamefont
  {\={O}nuki}, \citenamefont {Yamazaki}, \citenamefont {Omi}, \citenamefont
  {Ukon}, \citenamefont {Kobori},\ and\ \citenamefont
  {Komatsubara}}]{Onuki_1989}%
  \BibitemOpen
  \bibfield  {author} {\bibinfo {author} {\bibfnamefont {Y.}~\bibnamefont
  {\={O}nuki}}, \bibinfo {author} {\bibfnamefont {T.}~\bibnamefont {Yamazaki}},
  \bibinfo {author} {\bibfnamefont {T.}~\bibnamefont {Omi}}, \bibinfo {author}
  {\bibfnamefont {I.}~\bibnamefont {Ukon}}, \bibinfo {author} {\bibfnamefont
  {A.}~\bibnamefont {Kobori}}, \ and\ \bibinfo {author} {\bibfnamefont
  {T.}~\bibnamefont {Komatsubara}},\ }\href {\doibase 10.1143/JPSJ.58.2126}
  {\bibfield  {journal} {\bibinfo  {journal} {J. Phys. Soc. Jpn.}\ }\textbf
  {\bibinfo {volume} {58}},\ \bibinfo {pages} {2126} (\bibinfo {year}
  {1989})}\BibitemShut {NoStop}%
\bibitem [{\citenamefont {Araki}\ \emph {et~al.}(2011)\citenamefont {Araki},
  \citenamefont {Shiroyama}, \citenamefont {Ikeda}, \citenamefont {Kobayashi},
  \citenamefont {Seiro}, \citenamefont {Geibel},\ and\ \citenamefont
  {Steglich}}]{Araki_2011}%
  \BibitemOpen
  \bibfield  {author} {\bibinfo {author} {\bibfnamefont {S.}~\bibnamefont
  {Araki}}, \bibinfo {author} {\bibfnamefont {Y.}~\bibnamefont {Shiroyama}},
  \bibinfo {author} {\bibfnamefont {Y.}~\bibnamefont {Ikeda}}, \bibinfo
  {author} {\bibfnamefont {T.~C.}\ \bibnamefont {Kobayashi}}, \bibinfo {author}
  {\bibfnamefont {S.}~\bibnamefont {Seiro}}, \bibinfo {author} {\bibfnamefont
  {C.}~\bibnamefont {Geibel}}, \ and\ \bibinfo {author} {\bibfnamefont
  {F.}~\bibnamefont {Steglich}},\ }\href {\doibase 10.1143/JPSJS.80SA.SA061}
  {\bibfield  {journal} {\bibinfo  {journal} {J. Phys. Soc. Jpn.}\ }\textbf
  {\bibinfo {volume} {80}},\ \bibinfo {pages} {SA061} (\bibinfo {year}
  {2011})}\BibitemShut {NoStop}%
\bibitem [{\citenamefont {Fert}\ and\ \citenamefont {Levy}(1987)}]{Fert_1987}%
  \BibitemOpen
  \bibfield  {author} {\bibinfo {author} {\bibfnamefont {A.}~\bibnamefont
  {Fert}}\ and\ \bibinfo {author} {\bibfnamefont {P.~M.}\ \bibnamefont
  {Levy}},\ }\href {\doibase 10.1103/PhysRevB.36.1907} {\bibfield  {journal}
  {\bibinfo  {journal} {Phys. Rev. B}\ }\textbf {\bibinfo {volume} {36}},\
  \bibinfo {pages} {1907} (\bibinfo {year} {1987})}\BibitemShut {NoStop}%
\bibitem [{\citenamefont {Aliev}\ \emph {et~al.}(1983)\citenamefont {Aliev},
  \citenamefont {Brandt}, \citenamefont {Moshchalkov},\ and\ \citenamefont
  {Chudinov}}]{Aliev_1983}%
  \BibitemOpen
  \bibfield  {author} {\bibinfo {author} {\bibfnamefont {F.}~\bibnamefont
  {Aliev}}, \bibinfo {author} {\bibfnamefont {N.}~\bibnamefont {Brandt}},
  \bibinfo {author} {\bibfnamefont {V.}~\bibnamefont {Moshchalkov}}, \ and\
  \bibinfo {author} {\bibfnamefont {S.}~\bibnamefont {Chudinov}},\ }\href
  {\doibase 10.1016/0038-1098(83)90636-1} {\bibfield  {journal} {\bibinfo
  {journal} {Solid State Commun.}\ }\textbf {\bibinfo {volume} {47}},\ \bibinfo
  {pages} {693 } (\bibinfo {year} {1983})}\BibitemShut {NoStop}%
\bibitem [{\citenamefont {Sparn}\ \emph {et~al.}(1985)\citenamefont {Sparn},
  \citenamefont {Lieke}, \citenamefont {Gottwick}, \citenamefont {Steglich},\
  and\ \citenamefont {Grewe}}]{Sparn_1985}%
  \BibitemOpen
  \bibfield  {author} {\bibinfo {author} {\bibfnamefont {G.}~\bibnamefont
  {Sparn}}, \bibinfo {author} {\bibfnamefont {W.}~\bibnamefont {Lieke}},
  \bibinfo {author} {\bibfnamefont {U.}~\bibnamefont {Gottwick}}, \bibinfo
  {author} {\bibfnamefont {F.}~\bibnamefont {Steglich}}, \ and\ \bibinfo
  {author} {\bibfnamefont {N.}~\bibnamefont {Grewe}},\ }\href {\doibase
  10.1016/0304-8853(85)90482-2} {\bibfield  {journal} {\bibinfo  {journal} {J.
  Magn. Magn. Mat.}\ }\textbf {\bibinfo {volume} {47\&48}},\ \bibinfo {pages}
  {521} (\bibinfo {year} {1985})}\BibitemShut {NoStop}%
\bibitem [{\citenamefont {Fischer}(1989)}]{Fischer_1989}%
  \BibitemOpen
  \bibfield  {author} {\bibinfo {author} {\bibfnamefont {K.~H.}\ \bibnamefont
  {Fischer}},\ }\href {http://dx.doi.org/10.1007/BF01321909} {\bibfield
  {journal} {\bibinfo  {journal} {Z. Phys. B Con. Mat.}\ }\textbf {\bibinfo
  {volume} {76}},\ \bibinfo {pages} {315} (\bibinfo {year} {1989})}\BibitemShut
  {NoStop}%
\bibitem [{\citenamefont {Link}\ \emph {et~al.}(1996)\citenamefont {Link},
  \citenamefont {Jaccard},\ and\ \citenamefont {Lejay}}]{Link_1996}%
  \BibitemOpen
  \bibfield  {author} {\bibinfo {author} {\bibfnamefont {P.}~\bibnamefont
  {Link}}, \bibinfo {author} {\bibfnamefont {D.}~\bibnamefont {Jaccard}}, \
  and\ \bibinfo {author} {\bibfnamefont {P.}~\bibnamefont {Lejay}},\ }\href
  {\doibase 10.1016/0921-4526(96)00262-1} {\bibfield  {journal} {\bibinfo
  {journal} {Physica B}\ }\textbf {\bibinfo {volume} {225}},\ \bibinfo {pages}
  {207} (\bibinfo {year} {1996})}\BibitemShut {NoStop}%
\bibitem [{\citenamefont {Zlati\ifmmode~\acute{c}\else \'{c}\fi{}}\ \emph
  {et~al.}(2003)\citenamefont {Zlati\ifmmode~\acute{c}\else \'{c}\fi{}},
  \citenamefont {Horvati\ifmmode~\acute{c}\else \'{c}\fi{}}, \citenamefont
  {Milat}, \citenamefont {Coqblin}, \citenamefont {Czycholl},\ and\
  \citenamefont {Grenzebach}}]{Zlatic_2003}%
  \BibitemOpen
  \bibfield  {author} {\bibinfo {author} {\bibfnamefont {V.}~\bibnamefont
  {Zlati\ifmmode~\acute{c}\else \'{c}\fi{}}}, \bibinfo {author} {\bibfnamefont
  {B.}~\bibnamefont {Horvati\ifmmode~\acute{c}\else \'{c}\fi{}}}, \bibinfo
  {author} {\bibfnamefont {I.}~\bibnamefont {Milat}}, \bibinfo {author}
  {\bibfnamefont {B.}~\bibnamefont {Coqblin}}, \bibinfo {author} {\bibfnamefont
  {G.}~\bibnamefont {Czycholl}}, \ and\ \bibinfo {author} {\bibfnamefont
  {C.}~\bibnamefont {Grenzebach}},\ }\href {\doibase
  10.1103/PhysRevB.68.104432} {\bibfield  {journal} {\bibinfo  {journal} {Phys.
  Rev. B}\ }\textbf {\bibinfo {volume} {68}},\ \bibinfo {pages} {104432}
  (\bibinfo {year} {2003})}\BibitemShut {NoStop}%
\bibitem [{\citenamefont {Behnia}\ \emph {et~al.}(2004)\citenamefont {Behnia},
  \citenamefont {Jaccard},\ and\ \citenamefont {Flouquet}}]{Kamran_2004}%
  \BibitemOpen
  \bibfield  {author} {\bibinfo {author} {\bibfnamefont {K.}~\bibnamefont
  {Behnia}}, \bibinfo {author} {\bibfnamefont {D.}~\bibnamefont {Jaccard}}, \
  and\ \bibinfo {author} {\bibfnamefont {J.}~\bibnamefont {Flouquet}},\ }\href
  {http://stacks.iop.org/0953-8984/16/i=28/a=037} {\bibfield  {journal}
  {\bibinfo  {journal} {J. Phys.: Condens. Matter}\ }\textbf {\bibinfo {volume}
  {16}},\ \bibinfo {pages} {5187} (\bibinfo {year} {2004})}\BibitemShut
  {NoStop}%
\bibitem [{\citenamefont {Oganesyan}\ and\ \citenamefont
  {Ussishkin}(2004)}]{Oganesyan_2004}%
  \BibitemOpen
  \bibfield  {author} {\bibinfo {author} {\bibfnamefont {V.}~\bibnamefont
  {Oganesyan}}\ and\ \bibinfo {author} {\bibfnamefont {I.}~\bibnamefont
  {Ussishkin}},\ }\href {\doibase 10.1103/PhysRevB.70.054503} {\bibfield
  {journal} {\bibinfo  {journal} {Phys. Rev. B}\ }\textbf {\bibinfo {volume}
  {70}},\ \bibinfo {pages} {054503} (\bibinfo {year} {2004})}\BibitemShut
  {NoStop}%
\bibitem [{\citenamefont {Behnia}(2009)}]{Kamran_2009}%
  \BibitemOpen
  \bibfield  {author} {\bibinfo {author} {\bibfnamefont {K.}~\bibnamefont
  {Behnia}},\ }\href {http://stacks.iop.org/0953-8984/21/i=11/a=113101}
  {\bibfield  {journal} {\bibinfo  {journal} {J. Phys.: Condens. Matter}\
  }\textbf {\bibinfo {volume} {21}},\ \bibinfo {pages} {113101} (\bibinfo
  {year} {2009})}\BibitemShut {NoStop}%
\bibitem [{\citenamefont {Sondheimer}(1948)}]{Sondheimer_1948}%
  \BibitemOpen
  \bibfield  {author} {\bibinfo {author} {\bibfnamefont {E.~H.}\ \bibnamefont
  {Sondheimer}},\ }\href {\doibase 10.1098/rspa.1948.0058} {\bibfield
  {journal} {\bibinfo  {journal} {P. Roy. Soc. Lond. A Mat.}\ }\textbf
  {\bibinfo {volume} {193}},\ \bibinfo {pages} {484} (\bibinfo {year}
  {1948})}\BibitemShut {NoStop}%
\bibitem [{\citenamefont {Bel}\ \emph {et~al.}(2004)\citenamefont {Bel},
  \citenamefont {Behnia}, \citenamefont {Nakajima}, \citenamefont {Izawa},
  \citenamefont {Matsuda}, \citenamefont {Shishido}, \citenamefont {Settai},\
  and\ \citenamefont {Onuki}}]{Bel_2004}%
  \BibitemOpen
  \bibfield  {author} {\bibinfo {author} {\bibfnamefont {R.}~\bibnamefont
  {Bel}}, \bibinfo {author} {\bibfnamefont {K.}~\bibnamefont {Behnia}},
  \bibinfo {author} {\bibfnamefont {Y.}~\bibnamefont {Nakajima}}, \bibinfo
  {author} {\bibfnamefont {K.}~\bibnamefont {Izawa}}, \bibinfo {author}
  {\bibfnamefont {Y.}~\bibnamefont {Matsuda}}, \bibinfo {author} {\bibfnamefont
  {H.}~\bibnamefont {Shishido}}, \bibinfo {author} {\bibfnamefont
  {R.}~\bibnamefont {Settai}}, \ and\ \bibinfo {author} {\bibfnamefont
  {Y.}~\bibnamefont {Onuki}},\ }\href {\doibase 10.1103/PhysRevLett.92.217002}
  {\bibfield  {journal} {\bibinfo  {journal} {Phys. Rev. Lett.}\ }\textbf
  {\bibinfo {volume} {92}},\ \bibinfo {pages} {217002} (\bibinfo {year}
  {2004})}\BibitemShut {NoStop}%
\bibitem [{\citenamefont {{Onose, Y.}}\ \emph {et~al.}(2007)\citenamefont
  {{Onose, Y.}}, \citenamefont {{Li, Lu}}, \citenamefont {{Petrovic, C.}},\
  and\ \citenamefont {{Ong, N. P.}}}]{Onose_2007}%
  \BibitemOpen
  \bibfield  {author} {\bibinfo {author} {\bibnamefont {{Onose, Y.}}}, \bibinfo
  {author} {\bibnamefont {{Li, Lu}}}, \bibinfo {author} {\bibnamefont
  {{Petrovic, C.}}}, \ and\ \bibinfo {author} {\bibnamefont {{Ong, N. P.}}},\
  }\href {\doibase 10.1209/0295-5075/79/17006} {\bibfield  {journal} {\bibinfo
  {journal} {Europhys. Lett.}\ }\textbf {\bibinfo {volume} {79}},\ \bibinfo
  {pages} {17006} (\bibinfo {year} {2007})}\BibitemShut {NoStop}%
\bibitem [{\citenamefont {Izawa}\ \emph {et~al.}(2007)\citenamefont {Izawa},
  \citenamefont {Behnia}, \citenamefont {Matsuda}, \citenamefont {Shishido},
  \citenamefont {Settai}, \citenamefont {Onuki},\ and\ \citenamefont
  {Flouquet}}]{Koichi_2007}%
  \BibitemOpen
  \bibfield  {author} {\bibinfo {author} {\bibfnamefont {K.}~\bibnamefont
  {Izawa}}, \bibinfo {author} {\bibfnamefont {K.}~\bibnamefont {Behnia}},
  \bibinfo {author} {\bibfnamefont {Y.}~\bibnamefont {Matsuda}}, \bibinfo
  {author} {\bibfnamefont {H.}~\bibnamefont {Shishido}}, \bibinfo {author}
  {\bibfnamefont {R.}~\bibnamefont {Settai}}, \bibinfo {author} {\bibfnamefont
  {Y.}~\bibnamefont {Onuki}}, \ and\ \bibinfo {author} {\bibfnamefont
  {J.}~\bibnamefont {Flouquet}},\ }\href {\doibase
  10.1103/PhysRevLett.99.147005} {\bibfield  {journal} {\bibinfo  {journal}
  {Phys. Rev. Lett.}\ }\textbf {\bibinfo {volume} {99}},\ \bibinfo {pages}
  {147005} (\bibinfo {year} {2007})}\BibitemShut {NoStop}%
\bibitem [{\citenamefont {Amato}\ \emph {et~al.}(1989)\citenamefont {Amato},
  \citenamefont {Jaccard}, \citenamefont {Sierro}, \citenamefont {Haen},
  \citenamefont {Lejay},\ and\ \citenamefont {Flouquet}}]{Amato_1989}%
  \BibitemOpen
  \bibfield  {author} {\bibinfo {author} {\bibfnamefont {A.}~\bibnamefont
  {Amato}}, \bibinfo {author} {\bibfnamefont {D.}~\bibnamefont {Jaccard}},
  \bibinfo {author} {\bibfnamefont {J.}~\bibnamefont {Sierro}}, \bibinfo
  {author} {\bibfnamefont {P.}~\bibnamefont {Haen}}, \bibinfo {author}
  {\bibfnamefont {P.}~\bibnamefont {Lejay}}, \ and\ \bibinfo {author}
  {\bibfnamefont {J.}~\bibnamefont {Flouquet}},\ }\href
  {http://dx.doi.org/10.1007/BF00681532} {\bibfield  {journal} {\bibinfo
  {journal} {J. Low Temp. Phys.}\ }\textbf {\bibinfo {volume} {77}},\ \bibinfo
  {pages} {195} (\bibinfo {year} {1989})}\BibitemShut {NoStop}%
\bibitem [{\citenamefont {Solomon}\ and\ \citenamefont
  {Otter}(1967)}]{Solomon_1967}%
  \BibitemOpen
  \bibfield  {author} {\bibinfo {author} {\bibfnamefont {P.~R.}\ \bibnamefont
  {Solomon}}\ and\ \bibinfo {author} {\bibfnamefont {F.~A.}\ \bibnamefont
  {Otter}},\ }\href {\doibase 10.1103/PhysRev.164.608} {\bibfield  {journal}
  {\bibinfo  {journal} {Phys. Rev.}\ }\textbf {\bibinfo {volume} {164}},\
  \bibinfo {pages} {608} (\bibinfo {year} {1967})}\BibitemShut {NoStop}%
\bibitem [{\citenamefont {Huebener}\ and\ \citenamefont
  {Seher}(1969)}]{Huebener_1969}%
  \BibitemOpen
  \bibfield  {author} {\bibinfo {author} {\bibfnamefont {R.~P.}\ \bibnamefont
  {Huebener}}\ and\ \bibinfo {author} {\bibfnamefont {A.}~\bibnamefont
  {Seher}},\ }\href {\doibase 10.1103/PhysRev.181.701} {\bibfield  {journal}
  {\bibinfo  {journal} {Phys. Rev.}\ }\textbf {\bibinfo {volume} {181}},\
  \bibinfo {pages} {701} (\bibinfo {year} {1969})}\BibitemShut {NoStop}%
\bibitem [{\citenamefont {Ri}\ \emph {et~al.}(1994)\citenamefont {Ri},
  \citenamefont {Gross}, \citenamefont {Gollnik}, \citenamefont {Beck},
  \citenamefont {Huebener}, \citenamefont {Wagner},\ and\ \citenamefont
  {Adrian}}]{Ri_1994}%
  \BibitemOpen
  \bibfield  {author} {\bibinfo {author} {\bibfnamefont {H.-C.}\ \bibnamefont
  {Ri}}, \bibinfo {author} {\bibfnamefont {R.}~\bibnamefont {Gross}}, \bibinfo
  {author} {\bibfnamefont {F.}~\bibnamefont {Gollnik}}, \bibinfo {author}
  {\bibfnamefont {A.}~\bibnamefont {Beck}}, \bibinfo {author} {\bibfnamefont
  {R.~P.}\ \bibnamefont {Huebener}}, \bibinfo {author} {\bibfnamefont
  {P.}~\bibnamefont {Wagner}}, \ and\ \bibinfo {author} {\bibfnamefont
  {H.}~\bibnamefont {Adrian}},\ }\href {\doibase 10.1103/PhysRevB.50.3312}
  {\bibfield  {journal} {\bibinfo  {journal} {Phys. Rev. B}\ }\textbf {\bibinfo
  {volume} {50}},\ \bibinfo {pages} {3312} (\bibinfo {year}
  {1994})}\BibitemShut {NoStop}%
\bibitem [{\citenamefont {Xu}\ \emph {et~al.}(2000)\citenamefont {Xu},
  \citenamefont {Ong}, \citenamefont {Wang}, \citenamefont {Kakeshita},\ and\
  \citenamefont {Uchida}}]{Ong_2000}%
  \BibitemOpen
  \bibfield  {author} {\bibinfo {author} {\bibfnamefont {Z.~A.}\ \bibnamefont
  {Xu}}, \bibinfo {author} {\bibfnamefont {N.~P.}\ \bibnamefont {Ong}},
  \bibinfo {author} {\bibfnamefont {Y.}~\bibnamefont {Wang}}, \bibinfo {author}
  {\bibfnamefont {T.}~\bibnamefont {Kakeshita}}, \ and\ \bibinfo {author}
  {\bibfnamefont {S.}~\bibnamefont {Uchida}},\ }\href@noop {} {\bibfield
  {journal} {\bibinfo  {journal} {Nature}\ }\textbf {\bibinfo {volume} {406}},\
  \bibinfo {pages} {486} (\bibinfo {year} {2000})}\BibitemShut {NoStop}%
\bibitem [{\citenamefont {Pourret}\ \emph {et~al.}(2006)\citenamefont
  {Pourret}, \citenamefont {Behnia}, \citenamefont {Kikuchi}, \citenamefont
  {Aoki}, \citenamefont {Sugawara},\ and\ \citenamefont {Sato}}]{Pourret_2006}%
  \BibitemOpen
  \bibfield  {author} {\bibinfo {author} {\bibfnamefont {A.}~\bibnamefont
  {Pourret}}, \bibinfo {author} {\bibfnamefont {K.}~\bibnamefont {Behnia}},
  \bibinfo {author} {\bibfnamefont {D.}~\bibnamefont {Kikuchi}}, \bibinfo
  {author} {\bibfnamefont {Y.}~\bibnamefont {Aoki}}, \bibinfo {author}
  {\bibfnamefont {H.}~\bibnamefont {Sugawara}}, \ and\ \bibinfo {author}
  {\bibfnamefont {H.}~\bibnamefont {Sato}},\ }\href {\doibase
  10.1103/PhysRevLett.96.176402} {\bibfield  {journal} {\bibinfo  {journal}
  {Phys. Rev. Lett.}\ }\textbf {\bibinfo {volume} {96}},\ \bibinfo {pages}
  {176402} (\bibinfo {year} {2006})}\BibitemShut {NoStop}%
\bibitem [{\citenamefont {Johannsen}\ \emph {et~al.}(2008)\citenamefont
  {Johannsen}, \citenamefont {S\"ullow}, \citenamefont {Sologubenko},
  \citenamefont {Lorenz},\ and\ \citenamefont {Mydosh}}]{Johannsen_2008}%
  \BibitemOpen
  \bibfield  {author} {\bibinfo {author} {\bibfnamefont {N.}~\bibnamefont
  {Johannsen}}, \bibinfo {author} {\bibfnamefont {S.}~\bibnamefont {S\"ullow}},
  \bibinfo {author} {\bibfnamefont {A.~V.}\ \bibnamefont {Sologubenko}},
  \bibinfo {author} {\bibfnamefont {T.}~\bibnamefont {Lorenz}}, \ and\ \bibinfo
  {author} {\bibfnamefont {J.~A.}\ \bibnamefont {Mydosh}},\ }\href {\doibase
  10.1103/PhysRevB.78.121103} {\bibfield  {journal} {\bibinfo  {journal} {Phys.
  Rev. B}\ }\textbf {\bibinfo {volume} {78}},\ \bibinfo {pages} {121103}
  (\bibinfo {year} {2008})}\BibitemShut {NoStop}%
\end{thebibliography}
%

\end{document}